\documentclass[twocolumn]{aastex631}

\begin{document}

\newcommand{\kms}{km s$^{-1}\;$}
\newcommand{\msun}{M_\odot}
\newcommand{\rsun}{R_\odot}
\newcommand{\Lsun}{L_\odot}
\newcommand{\teff}{T_{\rm eff}}

\title{Evolved Eclipsing Binaries and the Age of the Open Cluster NGC 752
\footnote{Based on observations made with with the Hobby-Eberly
    Telescope, which is a joint project of the University of Texas at
    Austin, the Pennsylvania State University, Stanford University,
    Ludwig-Maximilians-Universit\"{a}t M\"{u}nchen, and
    Georg-August-Universit\"{a}t G\"{o}ttingen}}

\author[0000-0003-4070-4881]{Eric L. Sandquist}
\affiliation{San Diego State University, Department of Astronomy, San
  Diego, CA, 92182 USA}

\author[0000-0002-6998-2327]{Andrew J. Buckner}
\affiliation{San Diego State University, Department of Astronomy, San
  Diego, CA, 92182 USA}

\author[0000-0003-0509-2656]{Matthew D. Shetrone}
\affiliation{University of California, Santa Cruz, UCO/Lick Observatory, 1156 High Street, Santa Cruz, CA 95064, USA}
\affiliation{University of Texas, McDonald Observatory, HC75 Box
  1337-L Fort Davis, TX, 79734 USA}

\author{Samuel C. Barden}
\affiliation{Leibniz-Institut f\"{u}r Astrophysik Potsdam, An der Sternwarte 16, 14482 Potsdam, Germany}

\author[0000-0002-3007-206X]{Catherine A. Pilachowski}
\affiliation{Department of Astronomy, Indiana University, Bloomington, IN 47405, USA}

\author{Constantine P. Deliyannis}
\affiliation{Department of Astronomy, Indiana University, Bloomington, IN 47405, USA}

\author{Dianne Harmer}
\affiliation{National Optical Astronomy Observatory, 950 N. Cherry Avenue, Tucson, AZ 85719, USA}

\author[0000-0002-7130-2757]{Robert Mathieu}
\affiliation{Department of Astronomy, University of Wisconsin-Madison, 475 North Charter Street, Madison, WI 53706, USA}

\author{S\o ren Meibom}
\affiliation{Harvard-Smithsonian Center for Astrophysics, Cambridge, MA 02138, USA}

\author{S\o ren Frandsen}
\affiliation{Stellar Astrophysics Centre, Department of Physics and Astronomy, Aarhus University,
Ny Munkegade 120, DK-8000 Aarhus C, Denmark}

\author[0000-0001-9647-2886]{Jerome A. Orosz}
\affiliation{San Diego State University, Department of Astronomy, San
  Diego, CA, 92182 USA}

\correspondingauthor{Eric L. Sandquist}
\email{esandquist@sdsu.edu}

\begin{abstract}
We present analyses of improved photometric and spectroscopic
observations for two detached eclipsing binaries at the turnoff of the open
cluster NGC 752: the 1.01 day binary DS And and the 15.53 d BD $+$37 410.
For DS And, we
find $M_1 = 1.692\pm0.004\pm0.010 \msun$, $R_1 = 2.185\pm0.004\pm0.008 \rsun$,
$M_2 = 1.184\pm0.001\pm0.003 \msun$, and $R_2 = 1.200\pm0.003\pm0.005 \rsun$. 
We either confirm or newly identify unusual characteristics of both stars in the binary: the primary star is found to be slightly hotter than the main sequence turn off and there is a more substantial discrepancy in its
luminosity compared to models (model luminosities are too large by
about 40\%), while the secondary star is oversized and cooler compared to other main sequence
stars in the same cluster. The evidence points to non-standard evolution for both stars, but most plausible 
paths cannot explain the low luminosity of the primary star. 

BD $+$37 410 only has one eclipse per cycle, but extensive spectroscopic observations and the {\it TESS} light curve constrain the stellar masses well: $M_1 = 1.717\pm0.011 \msun$ and $M_2 = 1.175\pm0.005 \msun$. The radius of the main sequence primary star near $2.9\rsun$ definitively requires large 
convective core overshooting ($> 0.2$ pressure scale heights) in models for its mass, and multiple lines of evidence point toward an age of
$1.61\pm0.03\pm0.05$ Gyr (statistical and systematic uncertainties). Because NGC 752 is currently undergoing the transition from non-degenerate to degenerate He ignition of its red clump stars, BD $+$37 410 A directly constrains the star mass where this transition occurs.
\end{abstract}

\section{Introduction}

Open star clusters have long been testing grounds for models of how
stars change with age. However, a critical missing piece of
information has been precisely measured star masses. Stellar ages must
always be measured via models, and without mass information we are at
the mercy of systematic errors resulting from inaccuracies in the
physics and chemical composition we encode in the models. This study
is part of a larger project to measure precise masses (as well as
other characteristics) for stars in open clusters in order to identify
these issues, and improve our models of stars and measurements of age.

The star cluster NGC 752 is a relatively nearby and moderately old object.
Notable previous estimates of NGC 752's age range from 1.34 Gyr \citep{agueros} to
1.58 Gyr \citep{barti}, and have generally been based on analysis
color-magnitude diagram information for member stars in various filter combinations.
While the uncertainties in quoted ages are typically $40-60$ Myr, the uncertainties are
undoubtedly underestimates because they do not include systematic
errors involving correlated errors in distance, reddening, and
color-$T_{\rm eff}$ relations, as well as uncertainties in the physics
used in the stellar models. New data, such as stellar masses and precise distances, will 
help eliminate these issues.

NGC 752 is an interesting target for stellar astrophysics in a few
other ways. The cluster appears to be old enough that its
brightest main sequence stars, including those at the turnoff, just
miss the red edge of the instability strip for $\delta$ Sct pulsating
stars. \citet{breger69} found no pulsators in one of the only studies
of short-period variability in the cluster. One $\gamma$ Dor star is
known (PLA 455; \citealt{smalley}), and because $\gamma$ Dor pulsation is generally found
on the faint, red side of the $\delta$ Sct instability strip, this supports
the idea that the cluster's turnoff is just redward of the edge. 
The cluster's age puts it in an interesting range for constraining
the spindown of single main sequence stars, and it may be useful for calibrating
gyrochronology \citep{barnes}. \citet{agueros} presented a first study of the
rotation periods of NGC 752 stars, although they presented results on a
relatively small number of stars well below the turnoff in the K and M
spectral types. Based on the magnitude extent of its red clump, NGC 752 appears to contain two different kinds of core He burning stars: ones that ignited He burning in non-degenerate gas, and ones that had a degenerate flash ignition \citep{girardi00}. Measurement of the masses of evolved stars in the cluster and the cluster age would constrain this important transition in stars.

For precision measurements of stellar masses, binary systems are unmatched. For example, \citet{ebreview}
tabulate systems with masses and radii measured to better than $\pm3$\% uncertainty. As luck would have it, there are at least two eclipsing binaries at or near the turnoff of NGC 752 that can provide masses for astrophysical inquiries in the cluster.
The short-period detached eclipsing binary DS And has been known since the observations of
\citet{alks61}, with major studies of the binary by \citet{sandm} and
\citet{milone19}. From the standpoint of stellar physics, the two
stars in this binary are interesting because they fall in a range
($\sim1.2-2.0 \msun$) in which a convective core is springing up and
increasing in size as mass increases due to increasing core
temperature and stronger CNO cycle nuclear energy release. Ad hoc
algorithms for convective core overshooting have been introduced to
explain unexpected characteristics of the stars. The issue is strongly
linked to age because the extent of the mixed core affects the amount
of fuel available for hydrogen burning on the main sequence. Without a
proper accounting of how the convective core develops as a function of
mass, we should expect there to be systematic error in derived ages.

Despite the short span since the last study by \citet{milone19}, we
present a new analysis of DS And, and show that we have significantly improved the precision of measurements
of the stars in the binary. We describe extensive new time-series
photometry of the binary along with new high signal-to-noise
spectroscopy and improved analysis of previous spectroscopy, utilize
the vast amount of photometry on the cluster to discuss spectral
energy distributions for the binary's stars, and present new modeling to derive the
characteristics of the orbits and of the stars themselves.

BD $+$37 410 was identified as a double-lined spectroscopic binary by \citet{daniel}, and as an eclipsing binary by \citet{pribulla9}. It is a brighter system than DS And, which probably identifies it as having a more evolved primary star. The binary period (15.534 d) is long enough that the two stars have probably evolved independently of each other, without enough tidal interaction to circularize the orbit \citep{mandm}. Radial velocities were published by \citet{pribulla10} and \citet{mmu}, and a spectroscopic orbital solution was derived by \citeauthor{pribulla10} However, because the system only shows one eclipse per orbital cycle, it has not been studied in detail, as it is unlikely to provide reliably measured radii for the two stars. Despite this, the masses of the stars can be measured, and can provide additional limits on the cluster age because the primary star resides near the very tip of the main sequence.

\section{Cluster Properties}

Before discussing the observation and analysis of the binaries,
we briefly summarize cluster information that will be needed for a
proper discussion of age indicators.

\subsection{Chemical Composition}

Knowledge of the chemical composition of cluster stars will affect our
interpretation of the characteristics of the binary stars. For that
reason, we summarize spectroscopic abundance studies of NGC 752 stars in
Table \ref{compostab} and briefly discuss the cluster's composition.

An important part of examining the literature abundances is to ensure
that the abundance scale is correct relative to the Sun.  \citet{lum}
set the $\log gf$ values for their lines based on measurements of a
solar reference spectrum taken at the National Solar Observatory
\citep{wallace} after comparing to \citet[A09]{A09} abundances, and
determined NGC 752 star Fe abundances relative to the same A09
abundances.  \citet{guo} examined the composition of twilight spectra,
and found [M/H]$_\odot = 0.004 \pm 0.015$. \citet{bocek} ran their
abundance analysis on the \citet{katlas} integrated solar flux atlas
to derive solar Fe abundance. \citet{mad} used a solar spectrum with
the abundance analysis tuned to return a target $A$(Fe) value for it.
\citet{reddy} determined solar abundance of Fe using an ATLAS9 model
solar spectrum. \citet{cp11} conducted an abundance analysis of an ESO
solar spectrum of Ganymede. \citet{sestito} used a solar spectrum from
the Moon and adjusted their oscillator strengths to produce log
$n$(Fe)$_\odot$ = 7.52. \citet{ht92} also adjusted their oscillator
strengths to produce a solar iron abundance 
for solar equivalent widths.

From the most recent spectroscopic measurements, there seems to be
some agreement that NGC 752 stars have solar or slightly sub-solar
iron abundances --- straight or weighted averages of the values in Table 1 give [Fe/H]$=-0.02$. For the most part, we will assume [Fe/H]$= 0$ for NGC 752 stars.
There does not appear to be evidence of a
systematic difference between RGB and MS stars, as might
be expected if diffusion played a role in modifying surface abundances.

Comparisons with stellar models require that the abundances relative
to the Sun be translated into a heavy-element mass fraction $Z$. This
in turn requires us to know the solar value, but this is still
significantly uncertain itself. Recent values range from $Z_\odot =
0.0122$ \citep{asp05} to 0.0153 \citep{caffau}, with $0.0139\pm0.0006$
a recent re-evaluation by \citet{A20}. Most recently, \citet{magg} extensively re-evaluated many aspects of solar abundance determinations and arrived at $Z =0.0177$, showing that the higher metal abundance brings solar models into greater agreement with helioseismology constraints. The isochrone models we will use most frequently in comparisons below will
use $Z = Z_\odot = 0.0152$, but we will look at consequences of other possible metallicities.

\begin{deluxetable}{lcclc}
\tablewidth{0pt}
\tabletypesize{\scriptsize}
\tablecaption{Spectroscopic Abundance Measurements of NGC 752 Stars}
\tablehead{\colhead{} & \colhead{} & \colhead{$N$} & \colhead{} & \colhead{Reference\tablenotemark{a}}}
\startdata
[Fe/H] & $0.00\pm0.06$ & 10 & RGB & 1\\
{[}Fe/H] & $-0.01\pm0.06$ & 23,6 & RGB,MS & 2\\
{[}M/H]  & $-0.032\pm0.037$ & 36 & MS & 3 \\
{[}Fe/H] & $-0.02\pm0.05$ & 10 & RGB & 4\\
{[}Fe/H] & $-0.063\pm0.014$ & 33 & solar-type MS & 5\\
{[}Fe/H]$_I$ & $-0.04\pm0.03$ & 4 & RGB & 6\\
{[}Fe/H]$_{II}$ & $-0.02\pm0.02$ & 4 & RGB & 6\\
{[}Fe/H] & $+0.08\pm0.04$ & 4 & RGB & 7 \\
{[}Fe/H] & $+0.013\pm0.009$ & 7 & MS & 8 \\
{[}Fe/H] & $-0.102\pm0.022$ & 8 & MS & 9\\
\enddata
\label{compostab}
\tablenotetext{a}{1: \citet{bocek20}. 2: \citet{lum}. 3: \citet{guo}. 4: \citet{bocek}. 5:
  \citet{mad}. 6: \citet{reddy}. 7: \citet{cp11}. 8:
  \citet{sestito,taylor}. 9: \citet{ht92,taylor}.}
\end{deluxetable}

\subsection{Reddening and Distance}\label{reddist}

\citet{taylor} critically examined measurements of the cluster's
reddening, and settled on $E(B-V) = 0.044\pm0.0034$.
\citet{twarog} used Str\"{o}mgren $uvby$CaH$\beta$ photometry to derive
$E(B-V)=0.034\pm0.004$. Although these values disagree somewhat, the 
reddening seems to be quite small, and should not be a major uncertainty in the
analysis of cluster CMDs. We will generally use the larger Taylor value below.

The weighted average parallax for NGC 752 members in {\it Gaia} Data Release 2 (DR2)
has been determined as $2.2304\pm0.0027$ mas \citep{gaiacmd}, or
$2.239\pm0.005$ mas \citep{cg18}.  However, there are a number of studies
indicating that {\it Gaia} parallaxes are systematically offset to
smaller values (and implying that objects are systematically closer
than {\it Gaia} indicates; \citealt{lindegren,zinn}). 
\citet{bailer} determined distances of $431.7^{+3.8}_{-3.2}$ pc for DS And and $434.1^{+19.6}_{-18.5}$ pc for BD $+$37 410 from {\it Gaia} Early Data Release 3 (EDR3) data. The quoted uncertainties include consideration of uncertainties in the zero-point of the {\it Gaia} parallax scale. The corresponding distance moduli [$(m-M)_0 = 8.176^{+0.019}_{-0.016}$, and $8.19^{+0.11}_{-0.09}$] are in excellent agreement
with cluster determinations by \citet{agueros} and \citet{twarog}:
$8.21_{-0.03}^{+0.04}$ and $8.20\pm0.05$ (or around 436.5 and 438.5 pc), respectively.

These estimates are, however, very different than the distance determined by
\citet{milone19} as part of their binary star modeling of DS And: $(m-M)_0 = 8.390\pm0.018\pm0.060$, or
$477\pm4\pm12$ pc. About half of the discrepancy appears to be due to their allowance for a third light source in the system.

\subsection{Cluster Membership, Single Stars, and Spectral Energy Distributions}

\begin{figure*}
\plotone{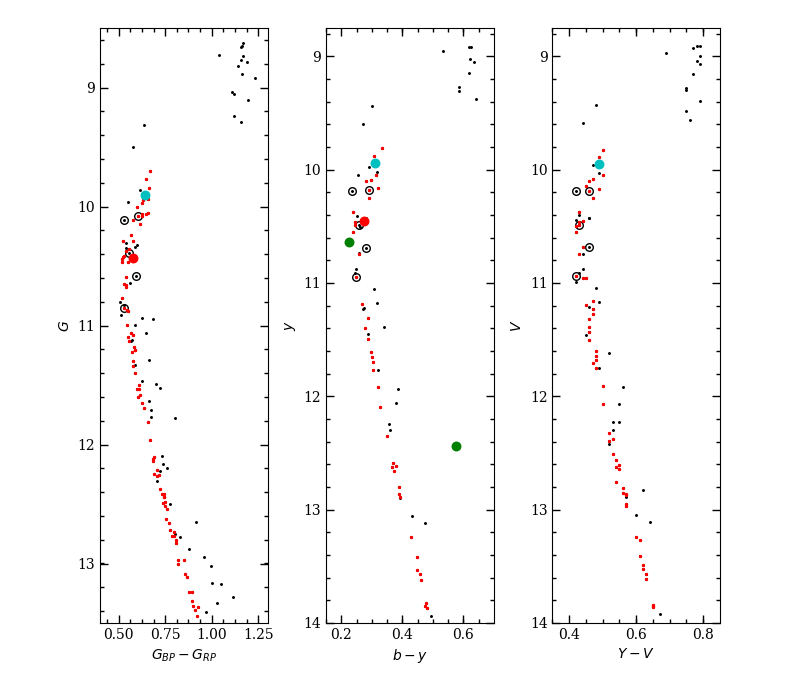}
\caption{Precise CMDs used for selecting likely single-star members of
  the NGC 752 cluster, with Gaia EDR3, Str\"{o}mgren \citep{twarog}, and
  Vilnius \citep{zdana} photometry shown from left to
  right. Gaia-selected likely binary members are shown in black, with
  photometrically selected single-star members in red. Open circles show 
  known chemically peculiar stars \citep{garrison}. Photometry of BD $+$37 410
  is shown with the cyan point,
  the DS And system at quadrature is the larger red point, and the
  positions of the DS And primary and secondary stars (as inferred from light
  curves) are shown with green points.
\label{bestcmds}}
\end{figure*}

Clear identification of NGC 752 stars is important for understanding
the properties of the stars of DS And in context, and for constraining
physics that affects the evolution of the stars at the cluster
turnoff. The most comprehensive membership studies of the cluster have
been \citet{platais}, \citet{daniel}, \citet{agueros}, and
\citet{cg18}. The last of these was the first application of Gaia Data
Release 2 to the cluster as a whole, and is the source of proper
motion and parallax cuts we use here. We also used radial velocity information from
\citeauthor{daniel} and \citeauthor{agueros} to identify
binary stars and remove them from consideration as representatives of single stars. After these cuts,
we attempted to eliminate photometric binaries using color-magnitude
diagrams using the most precise photometry. Unresolved binary systems
containing main sequence stars are found in a band brighter and redder
than the sequence of single stars.  We primarily used the Gaia Early Data Release 3 (EDR3)
diagram in ($G_{BP}-G_{RP}$,$G$), a Str\"{o}mgren ($b-y$, $y$) diagram
\citep{twarog}, and a Vilnius ($Y-V$, $V$) diagram \citep{zdana}, as
shown in Fig. \ref{bestcmds}.

Thanks to its small distance, NGC 752 stars have high signal-to-noise photometric observations in
many wavelength bands from deep in the ultraviolet to deep in the infrared. These data can be used to generate
well-sampled spectral energy distributions (SEDs), and fit for effective temperature $\teff$ and bolometric flux $F_{bol}$. In combination with Gaia distance measurements, it is possible to put many stars in the Hertzsprung-Russell Diagram (HRD) reliably. The sources for the photometry we used are given in Appendix \ref{photapp}, along with some
notes on the datasets and conversion to fluxes.

\section{Observations and Data Reduction}

\subsection{Time-Series Photometry}

We made primary use of photometry from the {\it Transiting Exoplanet Survey
  Satellite} ({\it TESS}; \citealt{tess}) taken in sector 18 of the
northern hemisphere campaign. DS And was not a target observed at
two-minute cadence, but {\it TESS} records full-frame images (FFIs)
every 30 minutes, and we derived a light curve from the FFIs using the
software {\tt eleanor} \citep{eleanor}. The TESS observations cover a
period of almost 25 days, with a data gap of about 3 days roughly in
the middle. In addition, the initial 2.5 days of the observations
showed variations that could not be corrected and appeared to be
affected by instrumental systematics, so we did not include those data
in the analysis. DS And is a fairly bright and isolated target, and so
we used a simple circular aperture of 2.5 pixel radius for
the photometry. We found that the {\tt eleanor} light curves
that used corrections from the built-in cotrending analysis (removing trends shared with nearby stars on the frames)
showed poorer consistency from cycle to cycle, so we used the light curve
based on raw fluxes.

We also used {\it TESS} photometry for BD $+$37 410. Two eclipses were observed during the campaign, although one
was observed in the initial days when the data quality was poorer. Using the Quick Look Pipeline (QLP) light curve
\citep{qlp}, the systematics were corrected well enough that we decided to include that eclipse in
the light curve analysis. For the later analysis, we also included photometry taken around the time of the non-eclipsing
conjunction.

Our ground-based photometric observations of DS And were taken with the 1 m
telescope at Mount Laguna Observatory with the CCD 2005 camera. This
camera uses a 2048$\times$2048 CCD with a field of view approximately
$13\farcm5$ on a side. Observations were taken during the course of one
season in 2010 with the intention of fully covering eclipse phases in
$BVR_CI_C$ filters, along with sparser coverage of out-of-eclipse
phases. The nights of observation are given in Table \ref{photobs}.

We took additional observations at the 1 m telescope in $U$ during
2020 in order to constrain the spectral energy distribution for the
component stars. These observations utilized a newer camera (called
UltraCam) with a similar field of view but lower read noise.

\begin{deluxetable}{cccDl}
\label{photobs}
\tablecaption{Photometry Observations at the Mount Laguna 1m Telescope}
\tablewidth{0pt}
\tablehead{
\colhead{UT date} & \colhead{Filters} & \colhead{mJD Start\tablenotemark{a}} & \colhead{$N_{obs}$}
}
\startdata
2010 Aug 31 & $R_CI_C$ & 5441.0273 & 79 \\
2010 Sep 1 & $BV$ & 55441.9295 & 80 \\
2010 Sep 7 & $B$ & 55446.7323 & 274 \\
2010 Sep 8 & $R$ & 55447.7294 & 390 \\
2010 Sep 11 & $V$ & 55450.7273 & 325 \\
2010 Sep 12 & $I_C$ & 55451.7193 & 465 \\
2010 Sep 17 & $VR_CI_C$ & 55457.9116 & 138 \\
2010 Oct 11 & $V$ & 55480.9278 & 10 \\
2010 Oct 14 & $V$ & 55483.6378 & 353 \\
2010 Oct 16 & $V$ & 55485.9787 & 194 \\
2010 Oct 17 & $I_C$ & 55486.6284 & 658 \\
2010 Oct 23 & $BR_C$ & 55492.6349 & 177 \\
2010 Oct 24 & $BR_C$ & 55493.6398 & 309 \\
2010 Oct 26 & $BR_C$ & 55495.8442 & 156 \\
2010 Nov 5 & $BVI_C$ & 55505.5874 & 395 \\
2010 Nov 17 & $VR_CI_C$ & 55517.9113 & 40 \\
2010 Nov 30 & $B$ & 55530.7061 & 304 \\
2010 Dec 2 & $V$ & 55532.5626 & 352 \\
2010 Dec 4 & $R_C$ & 55534.5901 & 387 \\
2010 Dec 9 & $BR_C$ & 55539.5825 & 190 \\
2020 Jul 7 & $U$ & 59037.8834 & 60\\
2020 Jul 24 & $U$ & 59055.8443 & 64\\
2020 Aug 10 & $U$ & 59072.7945 & 87\\
2020 Sep 4 & $U$ & 59097.7292 & 120\\
\enddata 
\tablenotetext{a}{mJD = BJD - 2400000.}
\end{deluxetable}

The images were processed in IRAF\footnote{IRAF is distributed by the
  National Optical Astronomy Observatory, which is operated by the
  Association of Universities for Research in Astronomy (AURA) under a
  cooperative agreement with the National Science Foundation.} with
standard techniques to subtract the overscan region for each image and
a master bias image, and divide out a normalized master flat field. A
somewhat non-standard step was correction for a nonlinearity in the
CCD response during the 2010 observations that resulted from
improperly set readout amplifier voltages.  The correction (D. Leonard
2016, private communication) used was
$ADU_{cor} = ADU*[1.01353-0.11576*(ADU/32767)+ 0.0296378*(ADU/32767)^2]$.
This was applied to flat field and cluster images using the procedure
{\tt irlincor} before the master flat field image was created and
applied to the cluster images. The corrections significantly improved
the consistency of light curves that were taken under differing
airmass and atmospheric conditions because we generally adjusted
exposure times to maintain high photon counts for DS And.

Aperture photometry was conducted using the {\tt DAOPHOT} package
within IRAF.  We employed 12 apertures to photometer all stars on each
frame, and subsequently constructed a curve of growth using the
routine {\tt mkapfile}. The curve of growth was used to correct
magnitudes measured in the aperture with the highest signal-to-noise
ratio to that of a common large radius. We then used an ensemble
photometry technique \citep{sand03,honey} to adjust the zeropoints for
each image in order to minimize the median magnitude difference for
stars from observation to observation. This improves the fidelity of
the differential photometry, but we saw clear evidence of changes in
the brightness of DS And on some nights.  We corrected for these
variations by applying nightly shifts.  The biggest need for these
corrections was in $R_C$ band, including $+0.04$ mag to 2010 Sep. 17,
$-0.03$ mag to 2010 Oct. 23, $+0.025$ mag to 2010 Dec. 2, and $+0.03$
mag to 2010 Dec 9. Smaller shifts were applied to a few nights in $B$,
but corrections did not appear to be needed for $V$ and $I_C$ bands.

We incorporated $V$ photometry from two other sources in order to
extend the time baseline of eclipse observations. \citet{breger}
presented 51 photoelectric observations from 5 nights in 1967,
providing complete coverage of the primary
eclipse. \citeauthor{breger} quotes a standard error of less than 0.01
mag for their differential photometry relative to a comparison
star. We also employed data from the Northern Sky Variability Survey
(NSVS; \citealt{nsvs}). Although the NSVS data typically were limited
to a few observations during a night spaced by a few days, the long
term observations produced a well-sampled light curve. In total there
were 288 observations from NSVS over the course of 225 days between
HJD 2451397 (1999 Dec. 6) and 2451623 (2000 Mar. 19).

We also considered the $UBVR_CI_C$ photoelectric photometry dataset
discussed in detail by \citet{milone19}. These data were taken in the
1982-1983 observing season. Observations in $BVR_CI_C$ filters were
taken on a 41 cm telescope at the University of Calgary's Rothney
Astrophysical Observatory, and in all five filters at the 90 cm
telescope at McDonald Observatory and at the 60 cm telescope at Table
Mountain Observatory. We initially assigned different but uniform
uncertainties to observations taken in the same filter at the same
facility, but we adjusted these uncertainties to be statistically
consistent with the rms scatter around best fit models. It is with
these kinds of adjustments that we weight the quality of different
photometric datasets relative to each other.

Finally, we used light curves from the Wide Field
Infrared Explorer (WISE; \citealt{wise}) in the $W1$ and $W2$
bandpasses provided by \citet{chen18}. These are depicted in
Fig. \ref{wiselc}.

\begin{deluxetable}{cccc}
\label{tsphot}
\tablecaption{Time-Series Photometry used in Modeling DS And}
\tablewidth{0pt}
\tablehead{
\colhead{mJD\tablenotemark{a}} & \colhead{Filter}  & \colhead{mag.} & \colhead{$\sigma$}}
\startdata
59037.884249 & $U$ & 11.8810 & 0.0222 \\
59037.889944 & $U$ & 11.8581 & 0.0180 \\
59037.892097 & $U$ & 11.8472 & 0.0150 \\
59037.893509 & $U$ & 11.8592 & 0.0174 \\
59037.895905 & $U$ & 11.8546 & 0.0108 \\
59037.898417 & $U$ & 11.8462 & 0.0117 \\
59037.900616 & $U$ & 11.8447 & 0.0096 \\
\enddata 
\tablenotetext{a}{mJD = BJD - 2400000.}
\tablecomments{Table \ref{tsphot} is published in its entirety in the
machine-readable format. A portion is shown here for guidance regarding
its form and content.}
\end{deluxetable}

\subsection{Spectral Energy Distributions (SEDs)}

Because of the vast amount of photometry available for NGC 752 stars, photometric
spectral energy distributions can be of use in deriving characteristics of
single cluster stars, as well as components of the eclipsing binaries. Appendix \ref{photapp}
discusses the photometry we have assembled for NGC
752 cluster members from the ultraviolet to deep in the infrared, as well as
details of our procedure for fitting the SEDs.  
After selecting cluster member stars from Gaia parallax and
proper motion information, we rejected spectroscopic binaries that have been previously identified
in the literature (and tabulated by \citealt{agueros}).
We then identified likely single main sequence
stars that were consistently near the blue edge of the MS band in CMDs
with precise photometry --- stars  more than about 0.03 mag from the blue edge of
the main sequence in the {\it Gaia} $(G_{BP}-G_{RP}, G)$ CMD were also rejected as likely binaries.

For BD $+$37 410, the luminosity ratio for the stars appears to be small ($L_2 / L_1 = 0.11$; \citealt{pribulla10}),
and so a fit places a lower limit on the effective temperature of the brighter primary star.
For the DS And system, the brightness is continually varying,
although with relatively small variations ($\sim0.06$ mag) outside of
the eclipses.  The eclipses cover approximately one-third of an
orbital cycle. Because the majority of literature photometric measurements of DS And
were probably not taken at phases that would give a measure of the
average brightness, significant scatter will
be introduced by the binary's variability even if all of the
measurements were taken out of eclipse. We have corrected for this in wavelength bands where
we have light curve information, as described below.

\subsubsection{Photometric Deconvolution and Effective Temperatures of DS And}\label{sedsec}

Calibrated photometry of the DS And binary can in principle give us
constraints on the characteristics of the component stars, but this is
complicated by the continuous photometric variation due to the eclipses
and ellipsoidal variations due to the non-spherical shapes of the
stars. Measurements of the system at the orbital quadratures represent
the total brightness of the system best because we see the maximal
area of both stars.  With well-sampled measurements during the
secondary eclipse, we observe the complete blocking of the flux
contribution of the secondary star when the stars are end-on. This
provides a direct measurement of the brightness of the secondary star,
and the brightness during secondary eclipse also provides a solid
lower limit to the flux from the primary star. With a correction for
the ellipsoidal shape, we can recover the flux from the primary star.
Colors measured during the secondary eclipse should be the colors of
the primary star.

\citet{milone19} raise the possibility of third light --- flux
contributions from an additional star that is blended with those from
the binary, whether due to a bound member of the system or an
unassociated star. Calculations of the secondary star's photometry are
unaffected by any third light, but if present, it would remain after
the secondary star's light is subtracted from the system photometry.
Regardless, precise determination of the color-magnitude diagram (CMD)
positions of the secondary star and the system with the secondary
star's light subtracted (relative to other stars in the cluster) could
help identify potential third light effects or other abnormalities due
to the close interaction of the stars in the eclipsing binary. The
photometric characteristics of the system (best-fit maximum brightness
and secondary eclipse depth, as well as luminosity ratios) are given
in Table \ref{dmtab}, and details of the calculations are given below.

We first assembled well-sampled light curves from available sources
and modeled each individually in order to establish the maximum
brightness of the system and measure the luminosity ratio (secondary
divided by total) in each band. We utilized $UBVR_CI_C$ light curves
from our observations and those of \citet{sandm}, as well as the NSVS
$V$-band light curve.

\begin{figure}
\plotone{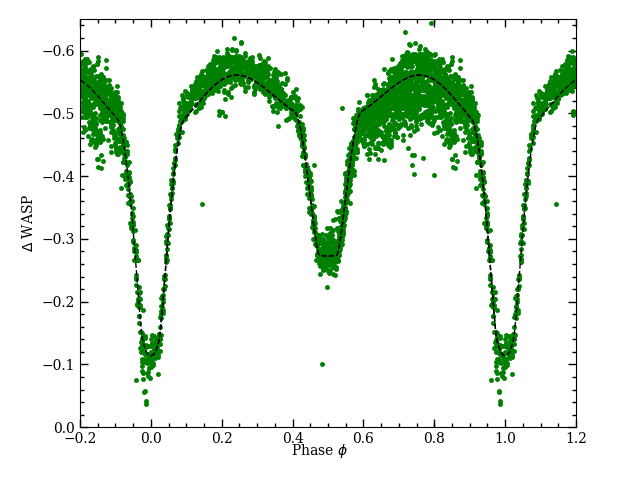}
\caption{DS And light curve from SuperWASP camera 141, computed
  differentially relative to nearby star NGC 752 165.\label{wasplc}}
\end{figure}

\begin{figure}
\plotone{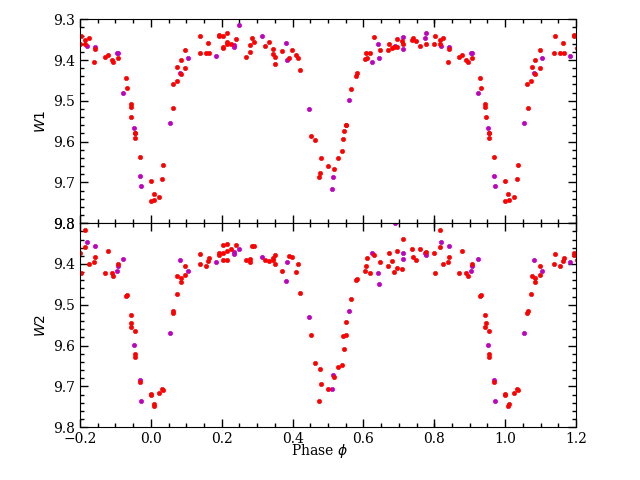}
\caption{DS And $W1$ and $W2$ light curves from WISE (magenta points) and NEOWISE-R (red)
  photometry \citep{chen18}.\label{wiselc}}
\end{figure}

The TESS light curve provides a measure of the relative brightness of
the two stars in a relatively broad bandpass centered near the $I_C$
filter. The filter for TESS is similar to the RP filter for Gaia, and
can provide guidance on decomposing the Gaia photometry of the two
stars in DS And. We also extracted a light curve from the SuperWASP
Data Release 1 database \citep{superwasp}, in part because the
SuperWASP filter (covering from around $400-700$ nm) is similar to the
Gaia BP filter. Because of systematics visible in the SuperWASP light
curves, we restricted ourselves to data from just one of the cameras
(ID number 141) and calculated variations differentially relative to
another nearby bright star (NGC 752 165). The light curve containing
3155 observations taken between 20 August and 17 December 2007, and a
model fit are shown in Figure \ref{wasplc}.

Finally, we utilized light curves in the WISE $W1$ and $W2$ bandpasses
\citep{chen18} in order to get information on the system in the far
infrared.

In cases where there are fewer observations at known times, the
magnitude measurements still contain information about the system flux
within the filter bandpass. With the well-determined ephemeris (such as a period uncertainty of
$2.0 \times 10^{-8}$ d), fits
to the observations allow us to estimate the total system flux
(measurable at the orbit quadratures) and the primary star
contribution using secondary star eclipse depths in similar
bandpasses. \citet{twarog} obtained limited time-series observations
of DS And in Str\"{o}mgren $uvby$ filters, typically 24 observations
in each filter mostly in the primary eclipse. These were provided by
B. Twarog (private communication).

The 2MASS infrared survey observed DS And at a single time (JD
2451127.7755), which was reasonably close to one of the quadratures at
phase 0.38. Because we have well-sampled lightcurves at wavelengths on
either side of the 2MASS bands, we can make informed estimates of
what the maximum magnitudes and luminosity ratios are.

GALEX observed DS And in the FUV passband at a phase close to the
quadrature ($\phi = 0.70$), and so we take this as being the system
magnitude.  The contribution from the cooler secondary star is
inferred to be less than 10\% in this wavelength range.

\begin{deluxetable}{ccccl}
\tablewidth{0pt}
\tabletypesize{\scriptsize}
\tablecaption{DS And Light Curve Characteristics}
\tablehead{\colhead{Filter} & \colhead{$m_{max}$} & \colhead{$\Delta m_\lambda$\tablenotemark{a}} & 
\colhead{$(L_2 /L_{tot})_\lambda$} & \colhead{Ref.\tablenotemark{b}}}
\startdata
FUV & $18.394\pm0.014$ &  &       &  \\
NUV & $14.254\pm0.011$ &  &       &  \\
$u$ & $12.153\pm0.006$ &  & 0.125 & 1\\
$U$ & $10.852\pm0.005$ &  & 0.077 & 2\\
$U$ & & 0.247 & 0.121 & 3\\
$v$ & $11.126\pm0.006$ &  & 0.124 & 1\\
$B$ & $10.833\pm0.009$ & 0.255 & 0.107 & 2\\
$B$ & & 0.251 & 0.125 & 3\\
$b$ & $10.726\pm0.007$ &  & 0.145 & 1 \\
$V$ & $10.439\pm0.004$ & 0.275 & 0.138 & 2\\
$V$ & & 0.265 & 0.148 & 3\\
$V$ & & 0.281 & 0.158 & 4\\
$y$ & $10.451\pm0.008$ &  & 0.161 & 1 \\
SuperWASP & 10.60 & 0.288 & 0.162 & 5 \\
$R_C$ & $10.222\pm0.005$ & 0.270 & 0.146 & 2\\
$R_C$ & & 0.300 & 0.167 & 3\\
$I_C$ & $9.959\pm0.012$ & 0.276 & 0.159 & 2\\
$I_C$ & & 0.287 & 0.172 & 3\\
TESS & & 0.2885 & 0.174 & \\
$J$ & 9.611 & & 0.194 & \\
$H$ & 9.441 & & 0.211 & \\
$K_s$ & 9.368 & & 0.213 & \\
$W1$ & 9.352 & 0.310 & 0.210 & 6 \\
$W2$ & 9.368 & 0.314 & 0.208 & 6 \\
$W3$ & 9.21 &  &  &  \\
\enddata
\tablenotetext{a}{Magnitude difference between brightness maximums at phase $\phi = 0.25$ 
and 0.75, and the brightness minimum in the secondary eclipse ($\phi = 0.5$).}
\tablenotetext{b}{1. \citet{twarog}. 2. \citet{milone19}. 3. This paper. 4. NSVS; \citet{nsvs}. 5. \citet{superwasp}. 6. \citet{chen18}.}
\label{dmtab}
\end{deluxetable}

Using fitted light curves, we can identify where the primary star
(plus possible third star) would be in some color-magnitude diagrams
(CMDs). The color during secondary eclipse should be representative of
the color of the primary star, and the flux during secondary eclipse
should be a lower limit to the flux of the primary star, only off by a
few percent due to ellipsoidal variations. We correct for the
ellipsoidal variation in each band using our best-fit models (see section
\ref{binary}).

\begin{figure}
\plotone{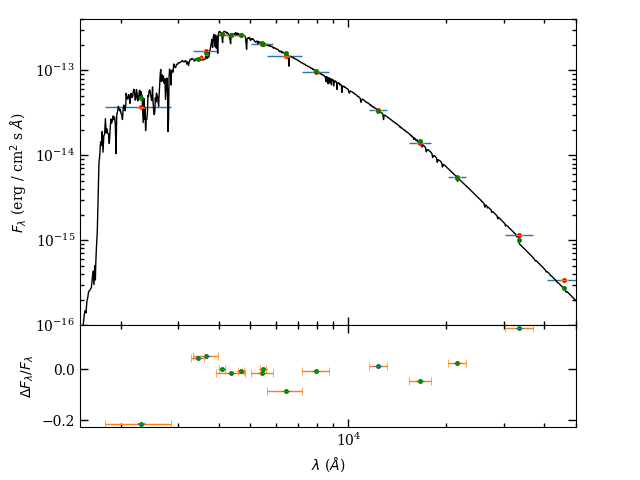}
\caption{Spectral energy distribution for DS And A as derived from
  light curves with observations of the secondary star eclipse (red
  points for photometry). A fitted ATLAS9 model for $T_{eff} = 7070$
  K, $\log g = 4.0$, and [Fe/H]=0 (solid line, and green points for
  integrations over filter response curves) is shown. Horizontal error
  bars represent the effective width of the filter.  {\it Bottom:}
  Fractional difference between the stellar fluxes and the best-fit
  model fluxes for the ATLAS9 (green) model.\label{ASED}}
\end{figure}

We have fitted the SED of DS And A using the algorithm discussed in
Appendix \ref{photapp}.
The primary star was best fit with 
$\teff = 7070$ K, as shown in Fig. \ref{ASED}.  We estimate the
statistical measurement uncertainty from the range in IRFM values
determined from the different 2MASS bands (40 K from minimum to
maximum). An uncertainty of 0.01 in $E(B-V)$ results in a shift of
approximately 50 K in the measured values, with lower reddening
leading to lower measured temperatures. We consider the reddening
uncertainty to be statistical here, and add it in quadrature with the
measurement uncertainty. We also assign an systematic uncertainty of $\pm100$ K to account for uncertainties in the model temperature scale.
Our final value is $7070\pm55\mbox{(stat.)}\pm100\mbox{(sys.)}$ K.
This compares well with $7056\pm21\pm140$ K from \citet{milone19},
found entirely through their modeling of different combinations of $UBVR_CI_C$ light curves.

The fitted bolometric flux at Earth for the primary star comes out
to be $F_{\rm bol} = 1.56 \times 10^{-9}$ erg s$^{-1}$ cm$^{-2}$.
Extinction uncertainties contribute at about a 3\% level in this
measurement, and uncertainties in the ultraviolet (where there is some
lack of coverage and mismatch between models and available
measurements, but also a relatively small 5\% contribution to the
total flux) result in another 1\% uncertainty. Together this produces
an uncertainty of $\pm0.05 \times 10^{-9}$ erg s$^{-1}$ cm$^{-2}$.

With the corrected {\it Gaia} distance (see section \ref{reddist}),
this produces $L_1 = 9.25 \pm 0.34 \Lsun$. This is in agreement with
the \citet{milone19} value $9.58 \pm 0.12 \pm 0.20 \Lsun$, with the difference
largely because they derived a larger distance modulus.

Comparisons to other NGC 752 stars are important for reliably
determining whether the properties of the stars in DS And have been
affected by their interactions within the short-period binary. While
the ($B-V$,$V$) photometry of \citet{daniel} is the only large dataset
of Johnson-Cousins photometry specifically of the cluster, there are
other sources of high signal-to-noise photometry that can be transformed into
the Johnson-Cousins system. We used relations from \citet{harmanec}
to convert Str\"{o}mgren $(b-y)$ and $(u-b)$ colors to $(B-V)$ and
$(U-B)$, and Geneva $U_G$, $B_G$, and $V_G$ magnitudes to Johnson $U$,
$B$, and $V$. We also used a relation from \citet{cousins} to
transform Str\"{o}mgren $(b-y)$ to $(V-I_C)$. For the Vilnius system,
we transformed from $(U_V-Y_V)$ to $(U-B)$ \citep{forbes}, and from
$(X_V-V_V)$ and $(Y_V-V_V)$ to $(B-V)$ \citep{straizys}.  We compared
these transformed photometric data to data taken directly in
Johnson-Cousins filters (APASS in $B-V$, and \citealt{taylorvri} in
$V-I_C$) where possible.

\begin{figure*}
\plotone{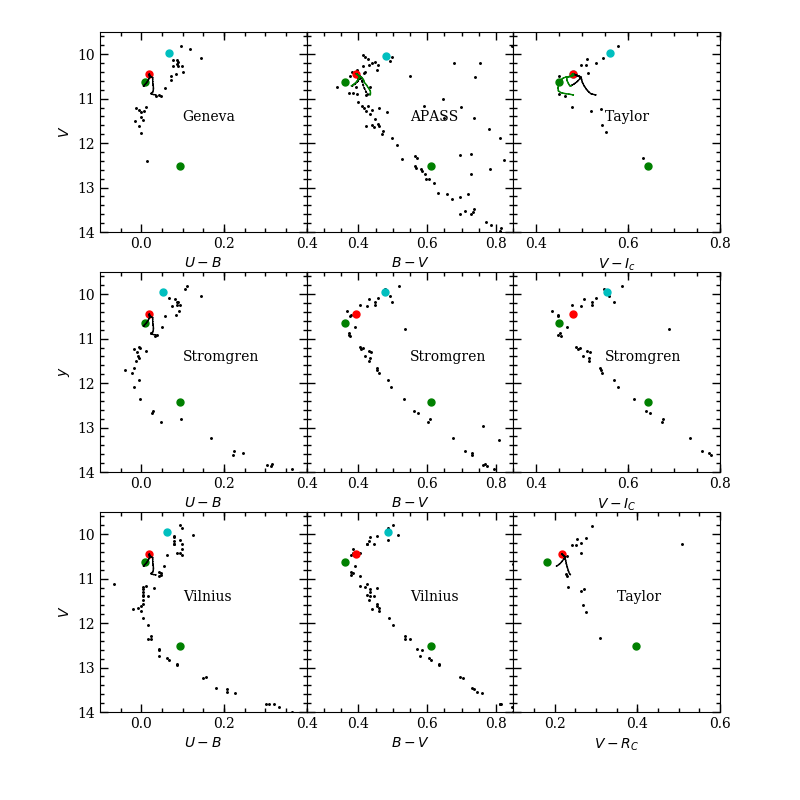}
\caption{Precise CMDs of likely single-star members of the NGC 752
  cluster.  Photometry of BD $+37$ 410 is shown with cyan points, the DS And system at quadrature is shown
  with red points, and the positions of the primary and secondary stars
  (as inferred from light curves) are shown with green points.
\label{cmds}}
\end{figure*}

As can be seen in the $(B-V)$ and $(V-I_C)$ CMDs in Figure \ref{cmds},
the primary star appears to have colors consistent with the main
sequence turnoff of the cluster, but it is significantly bluer in
$(U-B)$ than stars at the same $V$ magnitude. The bluest main sequence
stars in $(U-B)$ have $V \sim 11.7$, unlike the CMDs using other
colors, clearly illustrating effects of the Balmer jump on the
cluster's turnoff stars.  This implies that the primary star has a
weaker Balmer jump (more $U$ flux) than expected for similarly bright
stars.

Single stars at the cluster turnoff can provide an lower limit to the
effective temperature of the primary star. 
We fitted the photometric SEDs of the two bluest single-star
candidates in the Gaia $G_{BP}-G_{RP}$ color (BD+37 417/PLA 477, and
TYC 2829-1179-1), and find $\teff$ values of 6955 K and 6980 K.  These
two stars form a conservative bracket in brightness above and below DS
And A. The fitted bolometric fluxes were 1.69 and $1.25\times10^{-9}$
erg cm$^{-2}$ s$^{-1}$, which gives luminosities of 10.0 and
$7.42\Lsun$ using the Gaia distance.
The luminosities do indeed bracket the value for DS And A, and
these $\teff$ measurements are lower limits to that of DS And A if
there is no third light for DS And, and if interactions in the binary
have resulted in minor changes in surface temperature.

\begin{figure}
\plotone{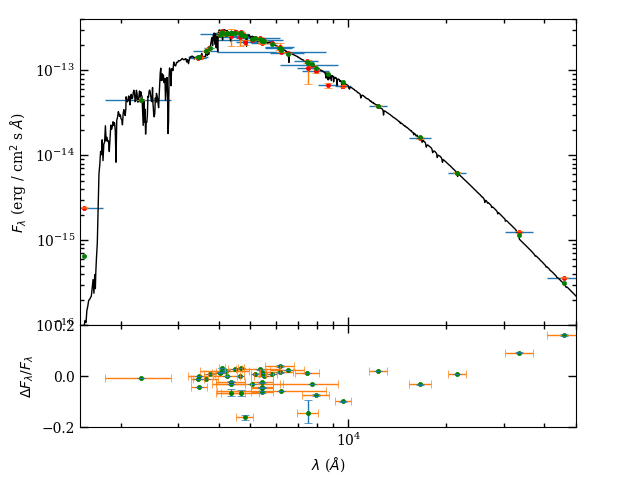}
\caption{{\it Top:} SED for turnoff star BD +37 417 (red points for
  photometry) and a fitted ATLAS9 model for $T_{eff} = 6955$ K, $\log
  g = 4.0$, and [Fe/H]=0 (solid line, and green points for
  integrations over filter response curves).  {\it Bottom:}
  Fractional difference between the stellar fluxes and the best-fit
  model fluxes for the ATLAS9 (green) model.\label{brightfig}}
\end{figure}

\subsubsection{The Secondary Star of DS And}\label{sec1}

The secondary star is consistently redder
or brighter than the locus of single main sequence stars. This fact is
based on the light lost during the total eclipse of the secondary, and
is a robust result of the light curve analysis,
but these characteristics are hard to explain. The secondary is very
unlikely to have evolved enough in size through single-star nuclear
evolution to reach this CMD position during the age of the
cluster. Light curve analysis below implies that the star is not
likely to be different from spherical by more than a few percent, and
the spectral analysis confirms that the star is rotating at or close
to synchronism with the orbit.  Short-period binaries with solar-type
stars frequently show evidence of inflation of radius, and this has
been attributed to inhibition of convection in stars with relatively
deep surface convection zones \citep{torres,clausen}. However, the
secondary star of DS And is massive enough that its surface convection
zone is likely to contain little of the star's mass.  In the 2.18 d
binary FL Lyr, the $0.96 \msun$ secondary star shows much clearer
signs of radius inflation than the primary, which is similar in mass
to DS And B (1.22 vs. $1.18 \msun$). In that case though, there is no
prior information about the binary's age, and so the primary star
could be mildly inflated or simply evolved, as its radius is larger
than expected for young main sequence stars. With a system of known age
like stars in this cluster, it should be possible to distinguish
between these scenarios.

\begin{figure}
\plotone{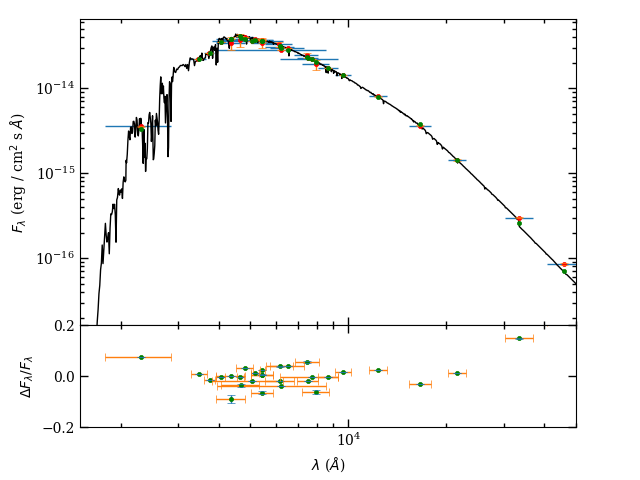}
\caption{{\it Top:} SED for PLA 255, the proxy for DS And B (red
  points) and a fitted ATLAS9 model for $T_{eff} = 6290$ K (solid
  line, and green points for integrations over filter response
  curves). Horizontal error bars represent the effective width of the
  filter. {\it Bottom:} Fractional difference between the stellar
  fluxes and the best-fit fluxes for the ATLAS9 (green) model.\label{faintpfig}}
\end{figure}

\begin{figure}
\plotone{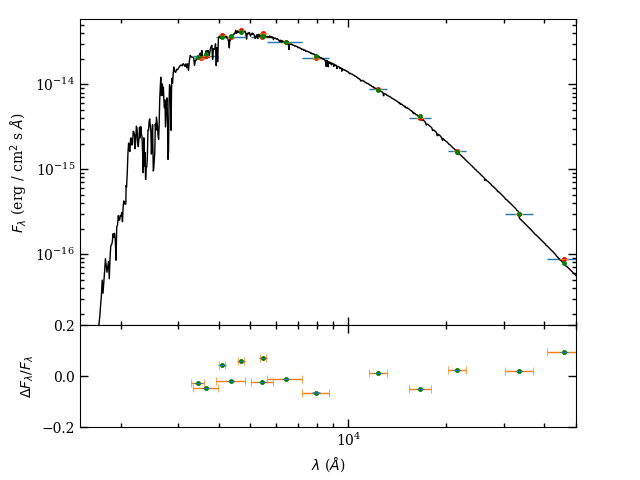}
\caption{{\it Top:} SED for DS And B (red
  points) and a fitted ATLAS9 model for $T_{eff} = 6100$ K (solid
  line, and green points for integrations over filter response
  curves). Horizontal error bars represent the effective width of the
  filter. {\it Bottom:} Fractional difference between the stellar
  fluxes and the best-fit fluxes for the ATLAS9 (green) model.\label{dsandbfig}}
\end{figure}

An examination of the CMDs in Figure \ref{cmds} implies that the color
deviations between DS And B and the main sequence are greater in
shorter-wavelength filters. This is borne out in comparisons with
stars at the same $V$ magnitude level, where the largest differences
are in the ultraviolet. PLA 255 is a good star for comparison, and a
fit to its SED returns $\teff = 6290$ K and $F_{bol} = 2.88 \times
10^{-10}$ erg cm$^{-2}$ s$^{-1}$ (see Fig. \ref{faintpfig}). A similar
fit to the 14 filter bands for which we can reconstruct the flux of DS
And B returns $\teff = 6100\pm43\pm100$ K and $F_{bol} = 2.95\pm0.07
\times 10^{-10}$ erg cm$^{-2}$ s$^{-1}$ (see Fig. \ref{dsandbfig}).
Reddening is accounted for in these fits, and contributions to the
temperature uncertainty from scatter in the IRFM calculations ($\pm25$
K) and reddening uncertainty ($\pm35$ K) are included.
Using the Gaia distance, we get a luminosity $L_2 = 1.75 \pm 0.05 \Lsun$.

Based on these comparisons, the secondary does appear to be
significantly lower in surface temperature than cluster main sequence
stars of the same brightness. We will discuss the secondary star in
more detail after the binary star analysis.

\subsubsection{BD $+$37 410}

As seen in Fig. \ref{cmds}, BD $+$37 410 lies comfortably among the most likely single stars brighter than the turnoff in 
optical and near-infrared CMDs. In the $U-B$ color, it is somewhat bluer than these stars. Regardless, this is one indication that any other bound companions or unassociated blended objects do not contribute greatly to the binary light. The measurement by \citet{pribulla10} of an optical luminosity ratio from spectral broadening functions ($L_2 / L_1 = 0.11$) localizes the secondary star to $G \approx 12.3$. If the star is a normal main sequence star of that brightness, it would have nearly the same color as the binary
and so a fit to the binary's SED places a lower limit on the effective temperature of the brighter primary star and an upper limit on its luminosity. 
The fit to the photometric SED with ATLAS9 models returns $\teff = 6480$ K, $\log g \approx 4.0$ (cgs), and $F_{bol} = 3.00 \times 10^{-9}$ erg s$^{-1}$ cm$^{-2}$. This puts an upper limit on the primary star $L_1 < 15.7 \Lsun$. 
Our spectroscopy below reveals a more complicated situation than \citeauthor{pribulla10} understood, and we will discuss this more.

\subsection{Spectroscopy}\label{specs}

\subsubsection{DS And}

For DS And, we obtained 17 spectra at the Hobby-Eberly Telescope (HET) with the
High Resolution Spectrograph (HRS; \citealt{tull}) between August 2010
and September 2011.  We used the configuration
HRS\_30k\_central\_600g5822\_2as\_2sky\_ISO\_GCO\_2x3 for a resolution
$R = 30,000$ on these rapidly rotating stars in all spectra. Exposure
times ranged from 1260 to 12240 s, and we only used the wavelength
range $4825-5760$ \AA ~ from the blue CCD in our subsequent
analysis. The data were reduced using the echelle package within IRAF
to remove the standard bias and scattered light, extract
one-dimensional spectra, and calibrate the wavelength scale.
Two observations were taken during morning twilight, and had
significant solar contamination, although this did not appear to
affect the measurement of the binary star velocities.

We also re-analyzed 15 observations of DS And taken using the Fibre-fed Echelle
Spectrograph (FIES; \citealt{fies}) on the 2.56 m Nordic Optical Telescope
(NOT) in August 2007, and July-November 2008. These spectra were taken
by Th. Mellergaard Amby, S. Frandsen, or NOT support staff. Exposures
were 1200 s (with one exception of 1800 s) taken using the medium
resolution fiber set-up. The spectra were processed using the standard
reduction pipeline
FIEStool\footnote{http://www.not.iac.es/instruments/fies/fiestool/FIEStool.html}.

\begin{figure}
\plotone{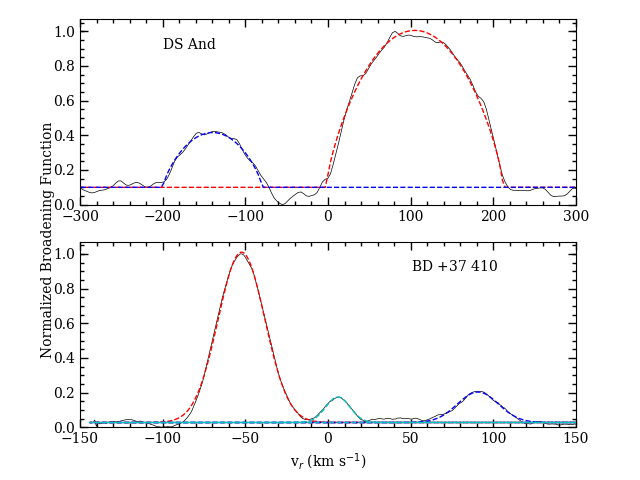}
\caption{Example broadening functions for the binaries. Primary star fits are in red, secondary star fits in blue, and tertiary star fits in cyan. {\it Top panel:} DS And broadening function from an HET HRS spectrum for HJD 2455811.801, with fits of rotational broadening functions with $v_{rot} \sin i = 106.6, 59.7$ km s$^{-1}$. {\it Bottom panel:} BD $+$37 410 broadening function from a WIYN Hydra spectrum for HJD 2452140.878, with Gaussian fits with $\sigma = 14.1, 12.4, 7.3$ km s$^{-1}$.
\label{bfsfig}}
\end{figure}

For the HRS spectra, we conducted a spectral disentangling
\citep{gonzo} to iteratively determine separated averaged spectra for
the two stars in the binary and their radial velocities at each
epoch. Figure \ref{bfsfig} shows the combined broadening function for one spectrum. 
Radial velocities were derived from fits of rotational
broadening profiles to the broadening functions \citep{rucin92}
derived from companion-subtracted spectra.  Synthetic spectral
templates from \citet{coelho} with temperatures of 6750 and 6000 K
were used for the primary and secondary star, respectively. We derived
our final radial velocities from a disentangling involving the entire
spectral range between 4825 and 5760 \AA.  In order to get an
empirical estimate of the velocity uncertainties, we
disentangled on three sections of the spectrum of roughly 300 \AA,
and computed the error in the mean.

FIES spectra cover wavelengths between 3700 and 7300 \AA ~ in 78
spectral orders, but the bluest and reddest orders were removed (due
to low signal-to-noise ratio and large sky contamination,
respectively). We attempted to find a middle ground between having
multiple measurements of the velocity from each spectrum (for the
purpose of understanding velocity uncertainties from the measurement
scatter) and signal in the broadening functions (from using larger
wavelength ranges to reduce noise). We settled on six sections of
approximately 300 \AA ~ each ($4060-4330$ \AA, $4329-4630$ \AA,
$4611-4952$ \AA, $4950-5270$ \AA, $5251-5523$ \AA, $5506-5860$ \AA),
generally including one strong line. Although these ranges overlap
somewhat, in all cases the sections were extracted from separate
orders of the spectrum. Orders were continuum normalized separately
before being merged together using the IRAF task scombine.
We fit rotational broadening profiles for both
stars using a modified version of the Python code
BF-rvplotter\footnote{https://github.com/mrawls/BF-rvplotter}. With
two exceptions the broadening functions for the two stars were
separately resolved.  One spectrum was removed from consideration
because the binary was in eclipse at the time of observation, and
reliable broadening function fits could not be obtained as a result.
In a second case, the broadening function of the secondary star
strongly overlapped that of the primary, and only the primary could be
reliably measured.

After comparison with the radial velocity measurements from HET HRS,
we found noticeable differences in the velocity curves. We re-examined
the FIES spectra and found that strong terrestrial lines (the [O I]
emission line at 557.33 nm, and the O$_2$ absorption band at 627.6 nm)
showed an offset from their expected values. We therefore
cross-correlated these orders against a spectrum from 2007 to derive a
correction for the velocities. The corrections ranged between $-2.3$ and
$-7.8$ km s$^{-1}$ for six spectra. With one exception, these spectra were taken between
phases of 0.1 and 0.4. Because these constituted a third of
the measurements used by \citet{milone19} in their analysis, we believe this resulted in 
a significant systematic error in their mass determinations.

The rotational broadening was determined for each star from the
broadening function fits, and we found these to be $105\pm0.7$ and
$58.3\pm0.8$ km s$^{-1}$, respectively for the primary and secondary stars. The quoted uncertainties are
errors in the mean of the rotational broadenings measured from each
spectrum. For comparison, \citet{milone19} found $106\pm3$ and $62\pm2$
km s$^{-1}$ for the two stars from the FIES spectra. As
\citeauthor{milone19} also found, these values are approximately consistent
with synchronous rotation for the binary. This is 27\% of breakup speed for the primary, and 13\% for the secondary.

\begin{figure*}
\plotone{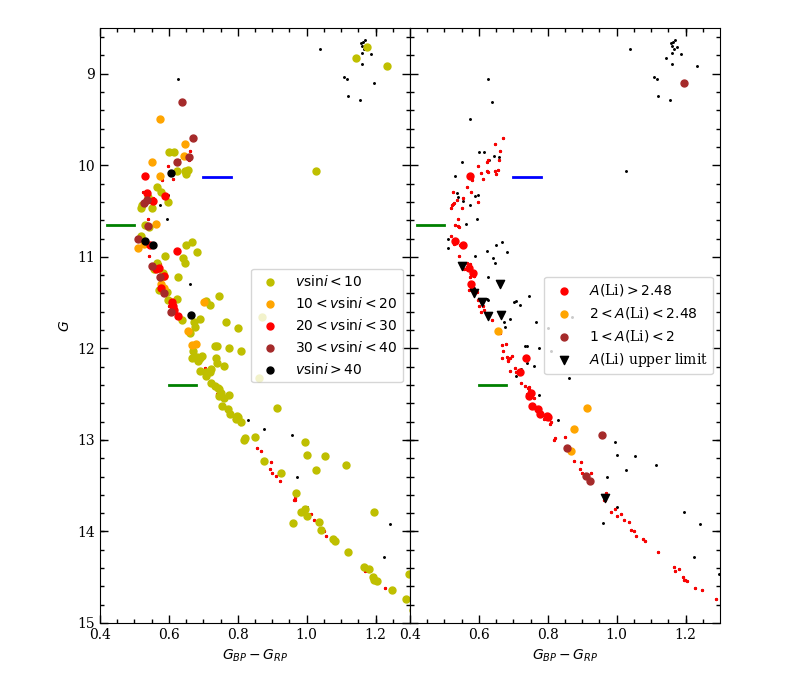}
\caption{Gaia CMDs with ({\it left panel}) rotational velocity and ({\it right panel}) Li abundance information.
Gaia-selected likely binary members are shown with small black points, and
  photometrically selected single-star members in red. Approximate $G$ magnitudes
  of the DS And components are shown with green lines, and BD $+$37 410 is shown with a blue line.
\label{rotmix}}
\end{figure*}

\begin{figure}
\plotone{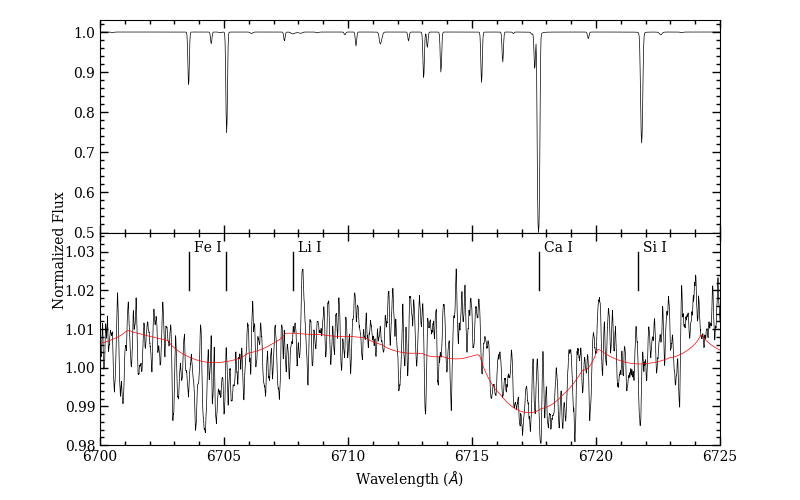}
\caption{{\it Top panel:} Synthetic spectrum for $\teff = 6750$ K from \citet{coelho}. {\it Bottom panel:} Disentangled
average spectrum of DS And A from NOT FIES observations in the region of the 6707.8 \AA ~ Li I line, with other strong lines identified. The rotationally-broadened synthetic spectrum (using $v_{rot} \sin i = 105$ km s$^{-1}$) is shown in red for comparison. 
\label{lispec}}
\end{figure}

As shown in Fig. \ref{rotmix}, the rotational velocities put these stars as comfortably rotating faster than other single stars at the same brightness level in the cluster \citep{mmu}, even accounting for inclination effects. The stars of DS And are at magnitude levels where the cluster main sequence stars have separate maximums in lithium abundance \citep{boesgaard}. Determination of Li abundances for the DS And stars are complicated by the rotational velocities, but the primary star at least appears to have had some Li depletion. The average spectrum of DS And A derived from a disentangling of the NOT FIES spectra is shown in Fig. \ref{lispec}. Fe, Ca, and Si lines are seen, but Li absorption is not obvious. This depletion is probably a sign of rotationally-driven mixing in the star, but further analysis is beyond the scope of the paper.

\subsubsection{BD $+37$ 410}

For BD $+$37 410, our primary source for velocity measurements was previously unpublished spectra taken with Hydra multi-fiber echelle spectrograph on the 3.5m  WIYN telescope at Kitt Peak National Observatory between 2001 Aug. 18 and 2009 Sep. 30. The spectra covered $4990-5250$ \AA ~ near the Mg I b triplet. Typically, $8-10$ exposures of around 1800 s were taken each night of observations, and had high signal-to-noise. After some experimentation, we settled on the $4990-5177$ \AA ~ range for our analysis as this appeared to provide the best signal on the fainter secondary star. We computed broadening functions from each spectrum using a synthetic template with $\teff = 6500$ K and $\log g = 4.0$ from \citet{coelho}, and fitted components with Gaussians. An example broadening function and fit is shown in Figure \ref{bfsfig}.

We were able to clearly detect the secondary star in all spectra with sufficient velocity separation, but in addition, we found
a third component with a radial velocity $5-7$ km s$^{-1}$, near the cluster mean and the binary systemic velocity $\gamma$. There are no other known sources likely to be contaminating the fiber aperture, so it is likely that the third object is associated with the binary, especially considering the large extent of the cluster on the sky. We can get a rough idea of the luminosity ratios for the stars from measurements of the broadening function areas if the temperatures of the components are similar. We find averages $A_2 / A_1 = 0.148 \pm0.003$ and $A_3 / A_1 = 0.088\pm0.002$ (error in the mean) from 44 observations. This implies that the secondary star contributes approximately 12\% of the system light in the optical, and the tertiary star contributes approximately 7\%. We will use this information to understand the eclipse depth in the analysis below, but it allows us to also identify a single cluster member star (PLA 641) as a rough proxy for the primary star based on its brightness relative to the binary.

We also made use of literature radial velocity measurements for the system. \citet{pribulla10} recorded 18 spectra with CfA Digital Speedometers \citep{ds}, reanalysed using two-dimensional cross correlation via TODCOR \citep{todcor}. \citet{mmu} recorded 32 spectra with the photoelectric scanner CORAVEL \citep{coravel1,coravel2} on the Swiss 1 m telescope at Observatoire de Haute-Provence.
Two additional measurements of the primary star were pulled from \citet{pribulla9}, who observed the system with the slit spectrograph at the 1.88m telescope of the David Dunlop Observatory. The radial velocity was measured from broadening functions derived from a comparison with an observed sharp-line spectral template. Finally, three velocity measurements were available from the LAMOST Data Release 4 \citep{tian}. Two of the the measurements with radial velocities near zero were clearly discrepant with the radial velocity curve, and were discarded. None of these velocity datasets have previously been analyzed together. We note that \citeauthor{mmu} measured the rotational speed of the primary star as $v_{rot} \sin i = 19.1 \pm 1.6$ \kms, in line with other upper main sequence stars.

\section{Analysis}

\subsection{Cluster Membership}

\citet{milone19} discussed the membership of DS And, and came to the
conclusion that the system is a likely member based on sky location,
CMD locations of the component stars, distance, extinction and
metallicity, radial velocity, and proper motion. Here we update a few
pieces of information related to membership.

We find that the systemic radial velocity $\gamma$ of DS And from both
spectroscopic datasets are fully consistent with recent mean cluster
velocity measurements. \citet{pila} found a mean velocity of
$4.9\pm0.7$ \kms from 6 stars. \citet{daniel} found a mean of
$5.5\pm0.6$ \kms from a larger sample of 33 stars. \citet{merm752}
found a mean of $4.68\pm0.11$ \kms from 15 cluster giants.
\citet{milone19} found a larger systemic velocity for DS And ($8.13\pm0.02\pm0.50$ km
s$^{-1}$) from velocities derived from the FIES spectra, and this was
consistent with older measurements of the cluster velocity
\citep{rebeirot,friel}. As we pointed out earlier, however, several of
these measurements were affected by zeropoint errors, and our newer
determination of $\gamma$ is much more consistent with the cluster
mean.

In proper motions, \citet{agueros} determined a membership probability
of 99.8\% by modeling field and cluster stars from ground-based
measurements. \citet{cg18} determined DS And to be a 100\% member from
{\it Gaia} Data Release 2 measurements of the clustering of proper
motion and parallax values.

For BD $+37$ 410, ground-based membership surveys \citep{francic,platais,daniel} have identified the star as a clear cluster member. With the availability of Gaia data, the membership is a little more ambiguous. While \citet{agueros} identifed the star as a cluster member, \citet{cg18} did not, most likely due to a proper motion in the right ascension direction ($\mu_{\alpha^*} = 10.99\pm0.12$ mas yr$^{-1}$) that is slightly larger than for most members (the cluster average is $\mu_{\alpha^*} = 9.8$ mas yr$^{-1}$; \citealt{gaiacmd}). \citet{bhatta} also did not identify the star as a possible member in their survey for stars in tidal tails, but the binary was outside of their search range in $\mu_{\alpha^*}$. \citet{boffin} identified the binary as a member in their wider search for tidal features of the cluster. BD $+$37 410 is only $0\fdg22$ from the cluster center on the sky, and deviations from cluster means in other quantities are small: the parallax ($2.28\pm0.12$ mas for the binary, versus 2.23 mas for the cluster), proper motion in the declination direction ($-12.17\pm0.11$ mas yr$^{-1}$, versus $-11.76$ mas yr$^{-1}$), and radial velocity ($4-6$ km s$^{-1}$ from our analysis, versus similar values for the cluster from the studies listed above). In addition, the binary's photometry puts it solidly among other cluster members at the bright end of the main sequence. As a result, we believe the evidence points clearly toward membership.

\subsection{Binary Star Modeling}\label{binary}

\subsubsection{DS And\label{dselc}}

We used the Eclipsing Light Curve code \citep[ELC;][]{elc} to simultaneously model the radial
velocities and multiband photometry for DS And.  We used a
differential evolution Markov Chain optimizer \citep{demc} for seeking
the overall best-fit model and exploring parameter space around that
model to generate a posterior probability sampling.  For the purposes of gauging some of the possible systematic errors, we conducted runs with different modelling runs approaches.

As a first model, we fitted the radial velocity data alone. Two of the parameters 
were the orbital period $P$, and a reference time of primary eclipse $t_c$.  Four
additional parameters mostly characterize the velocity variation: the
velocity semi-amplitude of the primary star $K_1$, mass ratio $q=M_2 /
M_1=K_1/K_2$, and systematic radial velocities\footnote{We allow for
  the possibility of differences for the two stars that could result
  from differences in convective blueshifts or gravitational
  redshifts.}  $\gamma_1$ and $\gamma_2$. We have also allowed for a
difference in zeropoint between the velocities from FIES and HRS
spectra. Previous examination of the times of secondary minimum has not 
revealed evidence of nonzero eccentricity \citep{sandm}. Because the binary has such a small orbital separation, tidal effects are assumed to maintain the system at zero eccentricity.

The subsequent model runs also utilized light curve data. These improve the precision of
measurements of period $P$ and reference time $t_c$, but require additional fitting parameters.
Because the stars orbit closely enough for tidal distortion of their
surfaces, we used Roche-lobe filling factors $f_1$ and $f_2$ in
modelling the stellar sizes and out-of-eclipse light variations. The
orbital inclination $i$ and temperature ratio $T_2 / T_1$ are primarily determined by eclipse data. The primary star temperature $T_1$ was included as a fitting parameter
as well, but is only weakly constrained by the light curves, and we consider the primary constraints on it to come from the SED fitting described earlier in \S \ref{sedsec}.

We included a parameter to account for potential contamination of the TESS light curve
by nearby stars. We find that this parameter is not constrained in the models, but we allow it
to vary in a range $0.05\pm0.05$ to account for this uncertainty in the fitting. For perspective, the TESS Input Catalog \citep[v. 8.2;][]{tic8} gives a contamination fraction of 0.0012 for DS And. We note that 
our models of the TESS light curves account for the long integration time in the full-frame images by integrating the computed light curves in each observed exposure window.

In two runs (labelled "albedo" in Table \ref{parmtab}), we experimented with fitting bolometric albedo coefficients $a_1$ and $a_2$ related to the reflection effect. The coefficient is expected to be 1 for stars with a radiative envelope, and 0.5 for stars with a convective envelope. The secondary star in DS And probably has a thin surface convection zone that we thought might affect the strength of the reflection effect. In fact, our fits point toward low values for the coefficients of both stars. The allowance for the lower coefficients seems to have the largest effect on the radius of the primary star, reducing it by about 0.6\%.

In two of our runs (with and without albedo coefficient fitting), we allowed limited fitting for limb darkening coefficients (labelled "LDC" in Table \ref{parmtab}).
In these runs, we fit for two quadratic limb darkening law
coefficients for each star in each filter band.  We used coefficients
in the form recommended by \citet{kipping} to produce a well-defined
area of parameter space that has physically realistic values where
each model star is forced to darken toward its limb with a
concave-down darkening curve.  If we allow the limb darkening
coefficients to fully float within the physically-realistic region,
however, the limb darkening coefficients take on values that are more
appropriate for stars with temperatures outside the range from our SED constraints, according to
theoretical models. As a result, we decided to constrain the
coefficients to small ranges near the expected values from \citet{claret13} and
\citet{claret18} based on the stellar temperatures inferred from SEDs.

Finally, we conducted runs where limb darkening coefficient fitting was replaced with angular-dependent intensity profiles derived from model atmospheres for the appropriate $\teff$ and $\log g$ (labelled "ATM" in Table \ref{parmtab}). While this utilizes information from detailed atmospheres, there still remains the possibility of systematic errors in those calculations.

The quality of the fitting is quantified by an overall $\chi^2$
derived from comparing the radial velocity and light curve data to the
models, as well as from how well an {\it a priori} constraint on the
primary star temperature (from SED fitting of single stars at the
cluster turnoff) was matched.  To try to ensure that different
datasets were given appropriate weights in the models, we empirically
scaled uncertainties on different datasets to produce a reduced
$\chi^2$ of 1 relative to a best-fit model when for the particular
dataset was fit on its own. For spectroscopic velocity measurements,
this meant scaling subsets of the data differently according to the
source spectrograph and the star (primary and secondary).

Approximate $1\sigma$ parameter uncertainties were derived from the
parts of the posterior distributions containing 68.2\% of the
remaining models from the Markov chains.  Gaussians were good
approximations to the posterior distributions for all parameters
except limb darkening coefficients and the inclination (which showed a small bimodality). 
The results of the parameter fits are provided in Table \ref{parmtab}. 

\begin{figure}
\epsscale{1.15}
\plotone{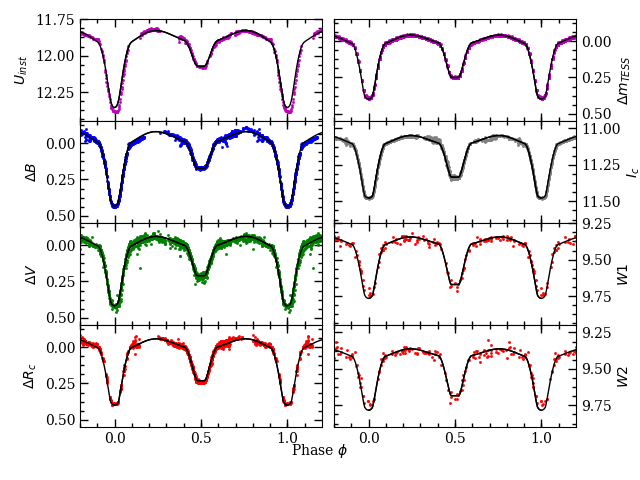}
\caption{Phased photometry for DS And compared with the best-fit binary
  models. $B$, $V$, $R_C$, and $TESS$ light curves are zeropointed to
  the magnitude at eclipse ingress/egress phases. $W1$ and $W2$
  photometry is in calibrated magnitudes. The $U$ curve is in
  instrumental magnitudes.
\label{phot}}
\end{figure}

\begin{figure}
\plotone{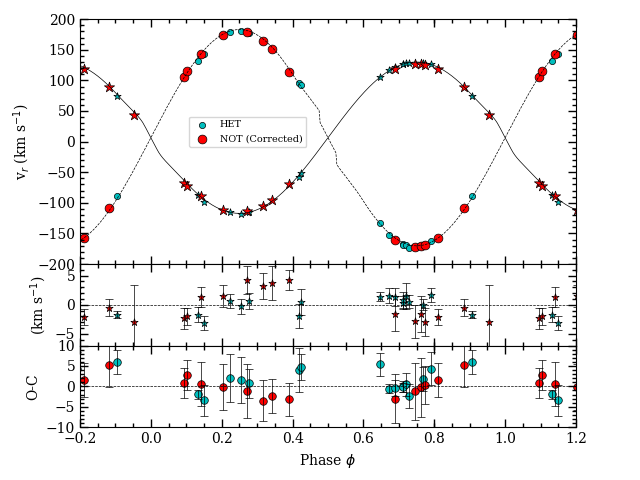}
\caption{{\it Top panel:} Radial velocities of the stars of DS And, phased to
  the orbital period, along with best fit models. {\it Bottom panels:}
$(O-C)$ residuals for the radial velocities.\label{rvplot}}
\end{figure}

A comparison of the light curves
with the best-fit model are shown in Figure \ref{phot}, and a
comparison of the radial velocity measurements with models are shown
in Figure \ref{rvplot}.
The fitting runs we present used photometry in 8 filter bands: $UBVR_CI_C$, TESS, and WISE $W1$ and $W2$ \citep{chen18}. The fits were generally very good with the exception of the primary eclipse depth for the bands at the extremes of the wavelength range ($U$ and $W2$).

\begin{deluxetable*}{lcccccc}
\tablewidth{0pt}
\tabletypesize{\scriptsize}
\tablecaption{Best-Fit Model Parameters for DS And}
\label{parmtab}
\tablehead{\colhead{Parameter} & \multicolumn{5}{c}{This Paper}  & \colhead{M19}\\
 & \multicolumn{4}{c}{RVs \& LCs} & \colhead{RVs Only} & \\
 & LDC & albedo & ATM & albedo}
\startdata
Constraints: & & \\
$T_1$ (K) & $7070\pm150$ \\
$T_2$ (K) & 6100 \\
\hline 
$\gamma$ (km s$^{-1}$) & & & & & & $8.13\pm0.02\pm0.50$\\
$\gamma_{1,NOT}$ & $4.63\pm0.03$ & $4.58\pm0.03$ & $4.54\pm0.03$ & $4.59\pm0.03$ & $5.25\pm0.33$ & \\
$\gamma_{1,HET}$ & $5.14\pm0.12$ & $5.32\pm0.12$ & $5.44\pm0.12$ & $5.29\pm0.12$ & $5.32\pm0.13$ & \\
$\gamma_{2,NOT}$ & $5.35\pm0.10$ & $5.22\pm0.10$ & $5.08\pm0.10$ & $5.21\pm0.10$ & $4.88\pm0.30$ & \\
$\gamma_{2,HET}$ & $3.77\pm0.28$ & $4.18\pm0.28$ & $4.59\pm0.28$ & $4.15\pm0.28$ & $3.52\pm0.26$ & \\
$P$ (d) & 1.01051956 & 1.01051955 & 1.01051956 & 1.01051955 & 1.0105217 & 1.010518870\\
$\sigma_P$ (d) & $2.0\times 10^{-8}$ & $2.0\times 10^{-8}$ & $2.0\times 10^{-8}$ & $2.0\times 10^{-8}$ & $1.4\times10^{-6}$ & $2.5\times10^{-8},1.20\times10^{-7}$\\
$t_c$ -2400000 & 55486.7694 & 55486.76941 & 55486.76941 & 55486.76942 & 55486.7676 & 36142.40281\\
$\sigma_{t_c}$ & 0.00004 & 0.00005 & 0.00004 & 0.00004 & 0.0006 & $0.00027,0.00114$\\
$q = M_2 / M_1$ & $0.705\pm0.003$ & $0.700\pm0.003$ & $0.699\pm0.003$ & $0.695\pm0.003$ & $0.707\pm0.003$ & $0.657\pm0.001$\\
$K_1$ (km s$^{-1}$) & $124.28\pm0.31$ & $123.86\pm0.32$ & $123.98\pm0.32$ & $123.47\pm0.32$ & $124.17\pm0.30$ & $119.75\pm0.98$\\
$K_2$ (km s$^{-1}$) & $182.00\pm0.52$ &                 &                 &                 & $175.58\pm0.59$ & $176.36\pm0.77$\\
$i$ ($\degr$) & $85.49^{+0.07}_{-0.11}$ & $85.99^{+0.12}_{-0.16}$ & $85.51^{+0.26}_{-0.09}$ & $86.18^{+0.08}_{-0.14}$ & & $89.35\pm0.19\pm0.45$\\
$f_1$ & $0.7425\pm0.0009$ & $0.7335\pm0.0015$ & $0.7417\pm0.0011$ & $0.7329\pm0.0014$ \\
$f_2$ & $0.4389\pm0.0010$ & $0.4395\pm0.0009$ & $0.4350\pm0.0012$ & $0.4355\pm0.0008$\\
$T_1$ (K) & $6900\pm10$ & $6771\pm10$ & $6855\pm16$ & $6770\pm7$\\
$T_2 / T_1$ & $0.8822\pm0.0006$ & $0.8881\pm0.0009$ & $0.8819\pm0.0005$ & $0.8869\pm0.0008$ \\
contam (TESS) & $0.05\pm0.05$ \\
$a_1$ & & $0.20\pm0.06$ & & $0.22\pm0.06$\\
$a_2$ & & $0.05^{+0.04}_{-0.03}$ & & $0.13\pm0.04$\\
\hline 
$M_1/\msun$ & $1.682\pm0.013$ & $1.688\pm0.013$ & $1.695\pm0.013$ & $1.702\pm0.013$ & & $1.655\pm0.003\pm0.030$\\
$M_2/\msun$ & $1.186\pm0.008$ & $1.181\pm0.008$ & $1.187\pm0.007$ & $1.183\pm0.007$ & & $1.087\pm0.005\pm0.040$\\
$R_1/\rsun$ & $2.188\pm0.006$ & $2.175\pm0.006$ & $2.196\pm0.006$ & $2.181\pm0.006$ & & $2.086\pm0.003\pm0.013$\\
$R_2/\rsun$ & $1.205\pm0.004$ & $1.205\pm0.004$ & $1.195\pm0.004$ & $1.195\pm0.003$ & & $1.255\pm0.005\pm0.012$\\
$\log g_1$ (cgs) & $3.984\pm0.0013$ & $3.990\pm0.0015$ & $3.985\pm0.0013$ & $3.992\pm0.0016$ & & \\
$\log g_2$ (cgs) & $4.351\pm0.0018$ & $4.349\pm0.0018$ & $4.358\pm0.0023$ & $4.357\pm0.0017$ & & \\
$L_1/\Lsun$ & $9.80\pm0.08$ & $9.02\pm0.08$ & $9.60\pm0.11$ & $9.05\pm0.06$ & & $9.58\pm0.12\pm0.20$\\ 
$L_2/\Lsun$ & $1.80\pm0.014$ & $1.72\pm0.013$ & $1.72\pm0.019$ & $1.68\pm0.012$ & & $1.77\pm0.03\pm0.06$\\
\enddata
\tablecomments{"LDC": limb-darkening coefficient fitting. "ATM": atmospheric model fitting. "albedo": fitting for reflection albedo coefficients for the two stars. See Section \ref{dselc} for parameter definitions.}
\end{deluxetable*}

The systematic differences for the fitted masses and radii from run to run generally exceed the statistical uncertainties, so that we consider both when assessing uncertainty. For the comparison, we use the mean of masses and radii from the four radial velocity and light curve fitting runs: $M_1/\msun = 1.692\pm0.004\pm0.010$, $R_1/\rsun=2.185\pm0.004\pm0.008$, $M_2/\msun = 1.184\pm0.001\pm0.003$, and $R_2/\rsun = 1.200\pm0.003\pm0.005$.
In these values, the uncertainties are statistical and systematic (indicated by run-to-run deviations), respectively.

\subsubsection{BD $+$37 410}

Modelling of this binary system utilized the same ELC program, but the fitted parameters
differed because of the configuration of the system. The stars were assumed to be spheres, in contrast to DS And. We fitted the radial velocity datasets and {\it TESS} lightcurve 
simultaneously with twelve parameters: orbital period $P$, reference time of periastron $t_0$, velocity semi-amplitude of the primary star $K_1$, mass ratio $q=M_2 /M_1=K_1/K_2$, eccentricity $e$, argument of periastron $\omega$, inclination $i$, the sum of 
the radii $(R_1 + R_2) / \rsun$, the radius ratio $R_1 / R_2$, the surface area ratio of the primary relative to the tertiary, and two quadratic limb darkening parameters for the eclipsed secondary star. The last three parameters relate to the modeling of the eclipse light curve but are poorly constrained by the data. We allow some variation around our best estimated values in order to allow those uncertainties to propagate into the uncertainties for other parameters.  We also fitted for systematic radial velocities $\gamma$ for each star in each dataset to allow for instrumental zeropoint differences and astrophysical effects (like differences in convective blueshifts or gravitational redshifts between the stars). 

We allowed for a small amount (1.52\%) of light contamination of the {\it TESS} aperture due to a few faint nearby stars. 
The value was taken from the {\it TESS} Input Catalog \citep[][v. 8.2]{tic}, and makes use of data from {\it Gaia} DR2.
As was the case with DS And, this parameter is not constrained in the models, but we allow it to vary in a range $0.015\pm0.005$ to account for this uncertainty in the fitting. The contamination has a minimal effect on the
fitted parameters, however. The "third light" contribution of the tertiary star seen in spectra is a more substantial effect.

The SED temperature measurement from the binary's combined light gives a lower limit to the temperature of the primary star ($T_1 > 6480$ K). We further estimated the CMD positions of the three stars assuming that they were all single-star cluster members with brightnesses consistent with our broadening function estimates from section \ref{specs}. Using SED fits to stars at those levels, we find $T_1 = 6620$ K, $T_2 = 6330$ K, and $T_3 = 6200$ K.\footnote{The template temperature used for the secondary star by \citet{pribulla10}, 6750 K, is unlikely to be correct. If the secondary star was that hot, it would likely require that the primary star reside off the cluster main sequence in order to match the observed photometry of the binary star. Their temperature for the primary star (6500 K) is, however, close to our determination.} These temperatures guided our selection of the ranges
the limb darkening coefficients were allowed to vary in, and the relative sizes of the primary and tertiary stars.

As a final note, we see evidence that primary star radial velocities near aphelion ($0.45 < \phi < 0.60$) showed systematic deviations of up to 5 \kms away from the best-fit model toward the tertiary star velocity, and in some cases, also the secondary star velocity. This seemed to be clear evidence that the primary star's velocity measurements were biased when the secondary star's lines weren't separately resolved. As a result, we left these measurements out of the final fitting because they had a demonstrable effect on the star masses.

\begin{figure}
\plotone{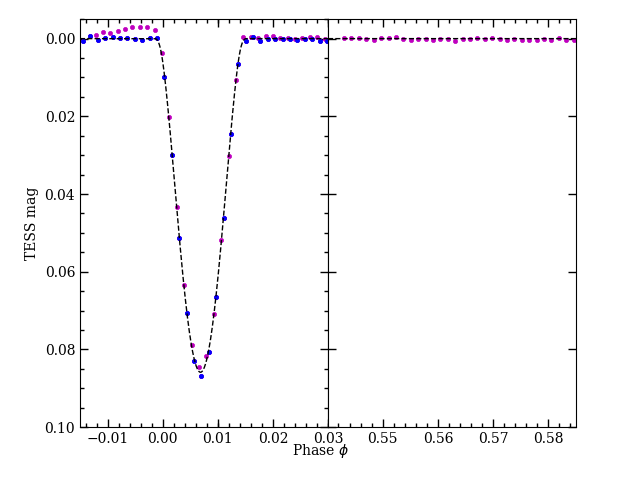}
\caption{Phased $TESS$ photometry for BD $+$37 410 compared with the best-fit binary
  models at the eclipse and non-eclipsing conjunction. The first $TESS$ eclipse is
  shown with magenta points, and the second with blue.
\label{bdphot}}
\end{figure}

\begin{figure}
\plotone{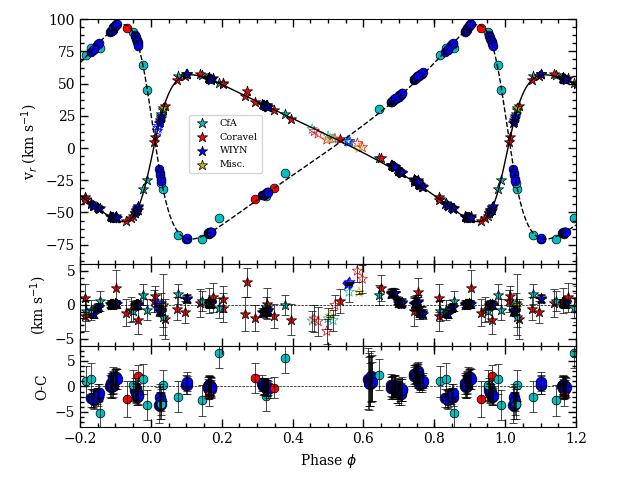}
\caption{{\it Top panel:} Radial velocities of the stars in BD $+$37 410, 
phased to the orbital period, along with best fit models. Empty symbols show 
measurements not used in the orbit fitting. {\it Bottom panels:}
$(O-C)$ residuals for the radial velocities.\label{rvplot_bd}}
\end{figure}

The results of the fitting are shown in Table \ref{parm2tab}, and model comparisons are shown in Figures \ref{bdphot} and \ref{rvplot_bd}. Although the system only has one partial eclipse per orbit, the inclination is well constrained. In a crude sense, the inclination is restricted to a range of values in which the eclipse occurs and is not total. Practically though, the eclipse reaches a depth that is close to what it would be for a total eclipse (approximately 12\%), so the light curve constrains the inclination to a much smaller range.

\begin{deluxetable}{lcc}
\tablewidth{0pt}
\tabletypesize{\scriptsize}
\label{parm2tab}
\tablecaption{Best-Fit Model Parameters for BD $+$37 410}
\tablehead{\colhead{Parameter} & \colhead{This Paper}  & \colhead{P10}}
\startdata
Constraints: & & \\
$T_1$ (K) & 6620 & 6500\\
$T_2$ (K) & 6330 & 6750\\
$T_3$ (K) & 5950 & \\
contam. (TESS) & $0.015\pm0.005$ \\
\hline 
$\gamma_{1,COR}$ & $5.96\pm0.04$ & \\
$\gamma_{1,CfA}$ & $4.75\pm0.05$ & $4.70\pm0.33$ \\
$\gamma_{1,WIYN}$ & $6.03\pm0.03$ & \\
$\gamma_{2,COR}$ & $3.83^{+0.09}_{-0.05}$ & \\
$\gamma_{2,CfA}$ & $5.96^{+0.07}_{-0.08}$ & \\
$\gamma_{2,WIYN}$ & $5.35\pm0.07$ & \\
$P$ (d) & 15.534837 & 15.53446 \\
$\sigma_P$ (d) & $^{+0.000007}_{-0.000006}$ & 0.00092 \\
$t_P -2400000$ & 45680.165 & 48491.979 \\
$\sigma_{t_P}$ & $^{+0.007}_{-0.008}$ & 0.037 \\
$q$ & $0.684\pm0.002$ & $0.689\pm0.020$ \\
$K_1$ (km s$^{-1}$) & $56.99\pm0.08$ & $57.03\pm0.54$ \\
$K_2$ (km s$^{-1}$) & $83.3\pm0.22$ & $82.8\pm1.5$\\
$e$ & $0.5142^{+0.0006}_{-0.0012}$ & $0.5156\pm0.0054$\\
$\omega$ ($\degr$) & $259.9\pm0.2$ & $260.6\pm1.3$\\
$i$ ($\degr$) & $81.7^{+0.07}_{-0.06}$ & $76-83$ \\
\hline 
$M_1/\msun \sin^3 i$ & $1.664\pm0.010$ & $1.638\pm0.076$\\
$M_2/\msun \sin^3 i$ & $1.139\pm0.005$ & $1.129\pm0.037$\\
$M_1/\msun$ & $1.717\pm0.011$ &  \\
$M_2/\msun$ & $1.175\pm0.005$ &  \\
$(R_1+R_2)/\rsun$ & $4.085\pm0.017$ & \\
$R_1/R_2$ & $2.44^{+0.16}_{-0.11}$ & \\
$R_1/\rsun$ & $2.899\pm0.013$ & \\
$R_2/\rsun$ & $1.186\pm0.005$ & \\
$\log g_1$ (cgs) & $3.748\pm0.004$ & \\
$\log g_2$ (cgs) & $4.360\pm0.003$ &  \\
\enddata
\end{deluxetable}

\subsection{Comparisons with Models}

\subsubsection{Luminosity and Temperature}

The calibrated photometry for DS And, along with measurements
of the depth of the total secondary eclipse, allowed us to identify
the contributions of both stars to the flux in many
wavebands in section \ref{sedsec}. In turn, this allowed us to fit the
SED and derive the effective temperatures and bolometric fluxes, as
well as luminosities in combination with the cluster
distance. For BD $+$37 410, the primary star is the dominant contributor to
the binary's light, making it easier to assess its characteristics directly from the SED. 

\begin{figure}
\plotone{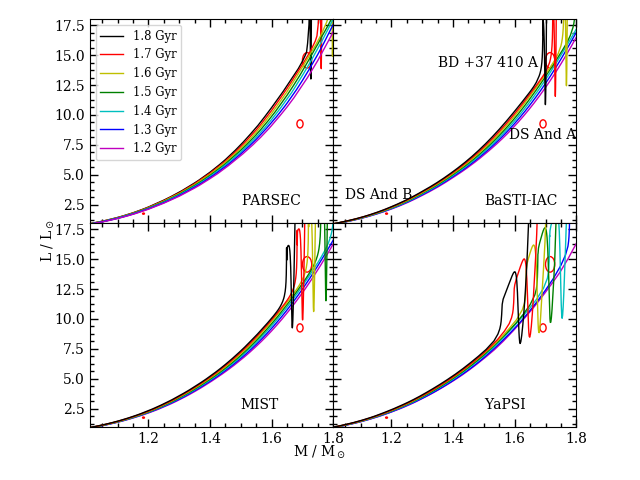}
\caption{Mass-luminosity plot for measured members of BD $+$37 410 and DS And, with
  $2\sigma$ uncertainties indicated by the red error ellipses. 
Models use $Z = 0.0152$, 
0.0149, 0.0142, and 0.0162,
  respectively for PARSEC \citep{parsec}, 
BaSTI-IAC \citep{basti_iac} isochrones.
MIST \citep{mist0,mist1}, and YaPSI
  \citep{yapsi} isochrones.
\label{toml}}
\end{figure}

Figure \ref{toml} shows a comparison of the components of these binaries with
isochrones in the mass-luminosity plane. With respect to BD $+$37 410 A, most models are ruled out because the model star leaves the turnoff and moves onto the subgiant branch
before it reaches the observed luminosity, in disagreement with the observations. This is a symptom
of error in the convective core overshooting parameters, and will be discussed in the next subsection.

A striking aspect of this
plot with regard to DS And is that the luminosity of the primary star is substantially lower
than the models for any reasonable age at the measured mass. \footnote{The measurements of \citet{milone19} put the primary closer to
the isochrones (due to lower mass and higher luminosity), but still
significantly subluminous compared to the models.} Even though DS And A appears to be at the turnoff for NGC 752, the star's mass-luminosity combination is
systematically off from the locus for turnoff stars predicted by different
isochrone sets. Model comparisons also indicate that DS And A is systematically cooler than 
the model predictions for a turnoff star of its mass, regardless of the isochrones used.
The amount of core overshooting has very little effect 
on these conclusions. Overshooting shifts turnoff masses and temperatures in such a way as to leave the relation the same, with only a change in the corresponding age for a given turnoff mass.
The resolution of the luminosity disagreement must be elsewhere.

\begin{figure}
\plotone{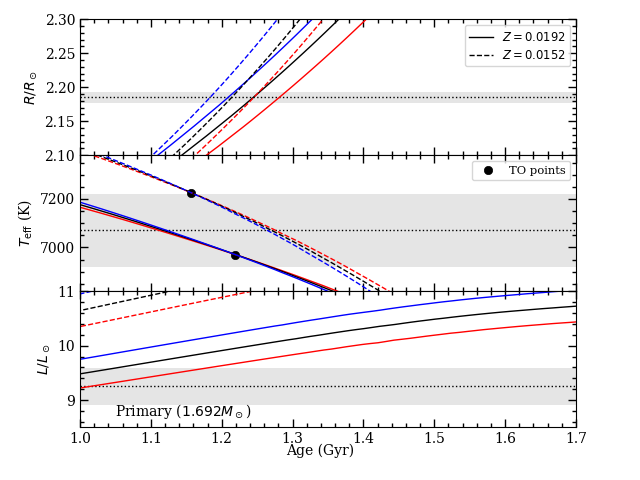}
\caption{Evolution of model characteristics for DS And A, according to
MIST \citep{mist0,mist1} models for two different metal contents $Z$.
Black lines show evolution tracks for the measured mass, with red and blue lines showing
tracks offset in mass by $\pm1 \sigma$. Black dots in the $\teff$ plot show the turnoff 
value for the corresponding $Z$. Grey bars show the $\pm 1\sigma$ uncertainty around the measured characteristics of DS And A.
\label{vst}}
\end{figure}

The temperature discrepancy could be solved and the luminosity discrepancy could be
alleviated (but not fixed) if the bulk metallicity of the cluster 
stars was higher, as shown in Figure \ref{vst}. As yet, there is no evidence that 
diffusion is affecting 
NGC 752 stars in such a way as to make their surface metal abundances appear
lower than the bulk metallicity --- main sequence and red giant stars
appear to have the same abundances \citep{lum}. But a higher solar metal
abundance $Z_\odot$ could produce a similar result without substantially affecting 
the radius comparisons in the previous section. 
As we will see below, higher metal content for cluster stars throws models out of agreement with
the observed main sequence.
We are left with the impression that DS And A's luminosity is 
systematically different than reasonable model predictions.

\begin{figure}
\plotone{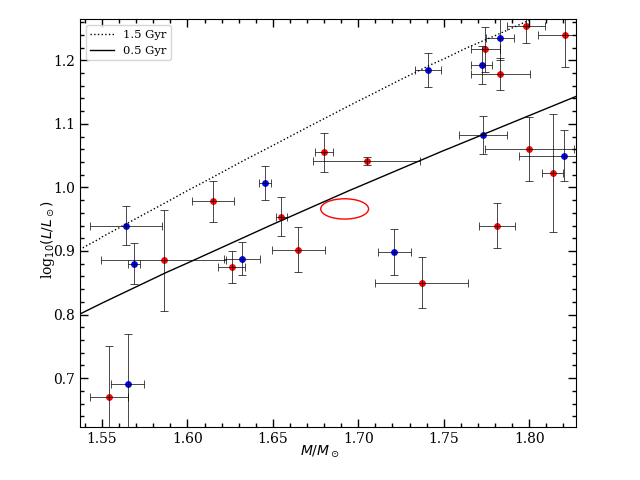}
\caption{Mass-luminosity plot for measured eclipsing binary stars
from the DEBCat database (retrieved 2019; \citealt{debcat}),
with $2\sigma$ error bars (showing primary stars in blue and secondary stars in red).
$2\sigma$ uncertainties for DS And A are indicated by the red error ellipse. 
PARSEC \citep{parsec} isochrones are shown for $Z = 0.0152$.
\label{mlstars}}
\end{figure}

This result rests in part on the fact that we can constrain the age
and/or evolutionary state of the star as well as its metallicity
because the binary is a member of the NGC 752 cluster. We can further test
the luminosity of the primary star by comparing with other
well-studied eclipsing binary stars of similar mass (see Table \ref{ecltab} and Figure \ref{mlstars}),
although these generally do not have supporting information that is as complete. 

As shown in the mass-luminosity comparison in Figure \ref{mlstars}, DS And A falls among stars of similar mass, 
but has a luminosity much smaller than predicted for a presumed cluster age of around 1.5 Gyr.
An HR Diagram emphasizes the unusual combination of characteristics for DS And A though --- while the other stars
in Table \ref{ecltab} fall near an evolution track for DS And A's mass, DS And A itself is significantly redder and/or less luminous. The stars that are likely to be most evolved (based on larger radius and lower temperature, like EI Cep B; \citealt{eicep}) have higher luminosities. 
The coolest of the stars in Table \ref{ecltab} (IO Aqr B) stands out, however. 
This star is in a relatively short-period orbit and is the largest of the stars in the table. \citet{ioaqr}
were able to model both stars in the binary with an isochrone of the same age, but only if they used 
an extremely super-solar metallicity [M/H]$=+0.3$ without spectroscopic evidence to justify this. This is 
potentially a symptom that the stars are less luminous than standard evolution models predict (and that the metallicity must be increased to compensate). If true, it may be another indication that stars in close binaries can
have lower-than-expected luminosities and effective temperatures, due in some way to their tidal interactions.

\begin{deluxetable*}{lcccccc}
\tablewidth{0pt}
\tabletypesize{\scriptsize}
\tablecaption{Eclipsing Binary Star Comparisons for DS And A}
\tablehead{\colhead{Star} & \colhead{$P$ (d)} & \colhead{$M/\msun$} & \colhead{$R/\rsun$} & \colhead{$\teff$ (K)} & \colhead{$L/\Lsun$} & Ref.\tablenotemark{a}}
\startdata
HW CMa A\tablenotemark{b,c} & 21.118 & $1.721\pm0.011$ & $1.643\pm0.018$ & $7560\pm150$ & $7.91\pm0.68$ & 1\\
AY Cam B & 2.735 & $1.705\pm0.036$ & $2.025\pm0.015$ & $7395\pm100$ & $11.01\pm0.16$ & 2\\
DS And A & 1.011 & $1.692\pm0.004$ & $2.185\pm0.004$ & $7070\pm150$ & $9.25\pm0.34$\\
EI Cep B\tablenotemark{b} & 8.439 & $1.6801\pm0.0062$ & $2.329\pm0.044$ & $6950\pm100$ & $11.35\pm0.81$ & 3\\
TV Nor B & 8.524 & $1.665\pm0.018$ & $1.550\pm0.014$ & $7800\pm100$ & $7.98\pm0.67$ & 4\\
IO Aqr B & 2.368 & $1.655\pm0.004$ & $2.493\pm0.017$ & $6336\pm125$ & $9.00\pm0.65$ & 5\\
V501 Mon A\tablenotemark{d} & 7.021 & $1.6455\pm0.0043$ & $1.888\pm0.029$ & $7510\pm100$ & $10.16\pm0.65$  & 6\\
\enddata
\label{ecltab}
\tablenotetext{a}{References: 1. \citet{hwcma}. 2. \citet{aycam}. 3. \citet{eicep}. 4. \citet{tvnor}. 5. \citet{ioaqr}. 6. \citet{v501mon}. 7. \citet{hoyman}}
\tablenotetext{b}{Am star}
\tablenotetext{c}{Measured [Fe/H]$=+0.33\pm0.15$.}
\tablenotetext{d}{Measured [Fe/H]$=+0.01\pm0.06$.}
\end{deluxetable*}

\begin{figure}
\plotone{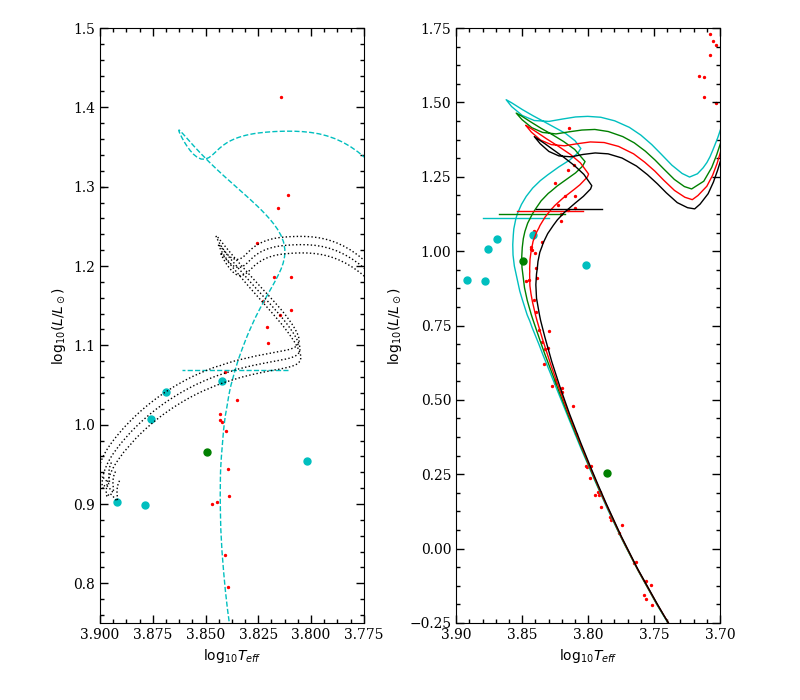}
\caption{HR Diagrams showing NGC 752 single star candidates ({\it red points}), DS And A ({\it green point}), and eclipsing binary components of similar mass ({\it cyan points}). {\it Left panel:} 1.4 Gyr isochrone ({\it cyan dotted line}) and a $1.692\pm0.01 \msun$ evolution tracks ({\it black dotted lines}) from the MIST database \citep{mist0,mist1} for $Z = 0.0152$. {\it Right panel:} PARSEC isochrones for $Z=0.0152$ and ages 1.4 (cyan), 1.5 (green), 1.6 (red), and 1.7 (black) Gyr.
\label{hrd}}
\end{figure}

\subsubsection{Mass and Radius}\label{mr}

High precision measurements of mass and radius from evolved stars in detached eclipsing binaries are capable of constraining age in a way that is independent of distance and reddening. The cases of the two binaries here are somewhat complicated, however. For DS And, there are questions about how strongly the
stellar characteristics have been influenced by star-star interactions, whether through the tidal interactions or through an unusual past history. For BD $+$37 410, the lack of a secondary eclipse means that the individual stellar radii are somewhat poorly measured, although the sum of the radii is constrained. In both cases it is helpful to have radius comparisons.

\begin{figure}
\plotone{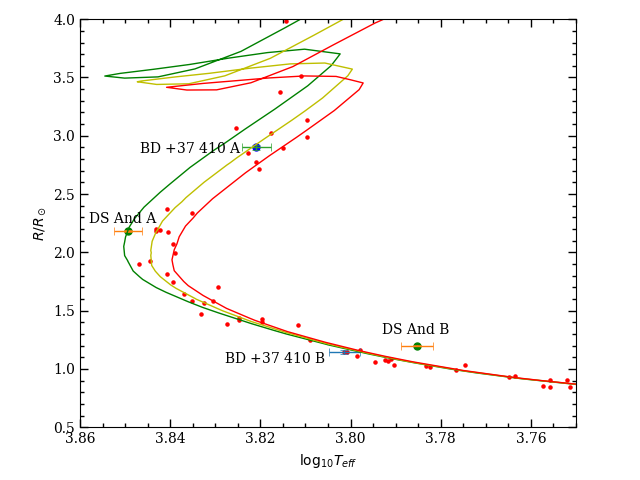}
\caption{Photometric radii versus effective temperature for single star candidates in NGC 752, compared to PARSEC isochrones for 1.5, 1.6, and 1.7 Gyr.\label{prvt}}
\end{figure}

Photometric radii can be derived from SED fits for bolometric flux and effective temperature, along with precise distances from {\it Gaia}. Calculations for our sample of single star candidates are shown in Figure \ref{prvt}. If we estimate the radius of the secondary star in BD $+$37 410 from stars at the appropriate brightness level, we have $R_2 = 1.14 \rsun$, which implies $R_1 = 2.92 \rsun$ when we subtract from the radius sum from the binary fitting.
These values are consistent with the results from binary fitting in Table \ref{parm2tab}.

Even with the uncertainties, BD $+$37 410 A is reliably seen to be a great deal larger than the primary star of DS And, and importantly, it has reached its present size while still on the main sequence. BD $+$37 410 A sits among a large number of other stars in CMDs like Fig. \ref{rotmix}, clearly showing that its evolutionary timescale is still slow and nuclear --- the cluster itself does not have any subgiants that are likely to be single stars because the evolution is too rapid there. While all reasonably isochrone models will have a star with the mass of BD $+$37 410 reach a radius of $2.9 \rsun$ at some point, the encoded physics in the isochrones determines whether it happens as a late main sequence star or as a rapidly-evolving subgiant. Therefore, the {\it combination} of observed mass, radius, and evolutionary state has the potential to {\it rule out} isochrone models that do not match.

\begin{figure*}
\plotone{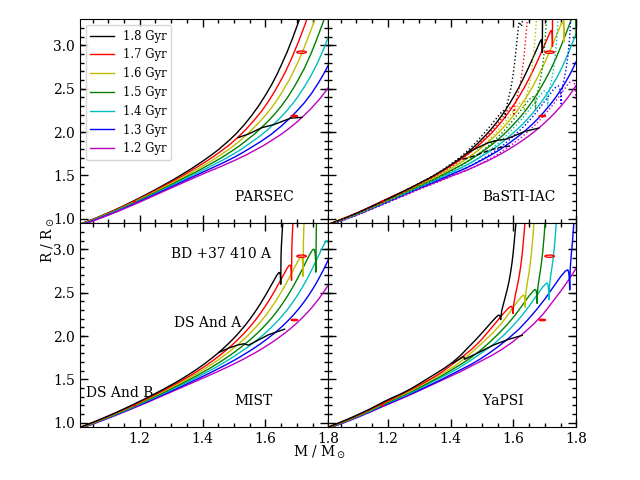}
\caption{Mass-radius plot for measured members of DS And and BD $+$37 410 A, with
  $2\sigma$ uncertainties indicated by the red error ellipses. 
Models use $Z = 0.0152$, 
0.0149, 0.0142, and 0.0162,
  respectively for PARSEC \citep{parsec},  
BaSTI-IAC \citep{basti_iac} isochrones.
MIST \citep{mist0,mist1}, and YaPSI
  \citep{yapsi} isochrones. In the BaSTI-IAC panel, dotted lines show
  models without convective core overshooting.
Solid black lines connect the lower turnoff points on the
different isochrones (with the dashed lines in the BaSTI-IAC panel for models without convective core overshooting).
\label{tomr}}
\end{figure*}

In Figure \ref{tomr}, we compare the characteristics of the BD $+$37 410 
primary and DS And stars with solar-metallicity isochrones. Metallicity has very little effect on the radii of model stars and is not the
cause of the differences between isochrone sets in the figure. Core overshooting primarily affects observable stellar characteristics at core 
hydrogen exhaustion and afterwards, and smaller amounts of overshooting allow the core to exhaust its core hydrogen supply earlier, before it has a chance to expand as greatly. The very last phase of the main sequence, as the star burns the last few percent of its core hydrogen, shows up as a short downward dip in the mass-radius isochrones.
Among the models we plot, the YaPSI models have the least overshooting in the age range we are examining ($\sim0.1H_P$ for stellar masses $1.6\msun < M < 1.8 \msun$, where $H_P$ is the pressure scale height). The PARSEC models have the largest overshooting ($\sim0.25H_P$ for $M > 1.5\msun$). 
PARSEC \citep{parsec} and BaSTI-IAC \citep{basti_iac} model stars reach a radius of $\sim3\rsun$ while on the main sequence, while models with less overshooting only hit that size after starting the subgiant branch. This fact is very important for age determination in NGC 752 because by ruling out models with smaller convective core overshooting it greatly reduces the systematic error associated with uncertainties in the overshooting and allows us to focus on models that realistically match observations. The mass-radius combination for BD $+$37 410 A points to an age of $1.61\pm0.03$ Gyr. The stellar radii are not very sensitive to metal content, but an increase in metallicity of 0.05 dex (due to observational error or uncertainties in the solar metal content) produces an larger age by 0.05 Gyr. We quote this source of uncertainty separately from the statistical uncertainties associated with the fitting of BD $+$37 410 A's radius.

It is important to emphasize that this conclusion would not have been possible without the mass measurements in particular. Reasonable fits to the CMDs by themselves are possible with all of the isochrone sets we have discussed here. However, the fits with isochrones having smaller amounts of convective overshooting have systematically lower ages. What we are showing here is that the models with the smallest amounts are not consistent with all of the available data, and they must be eliminated from consideration when determining the absolute age of the cluster.

\subsubsection{The Stars of DS And, Their History, and Their Future}

Evidence from earlier studies indicated that the characteristics of DS And B have been altered away from that of a normal single star, as it is found well to the red of the main sequence. In section \ref{sec1}, we derived characteristics of DS And B star from photometry and the Gaia distance. 
The temperature analysis indicated that it was approximately 200 K lower than stars at similar brightness. The mass-radius comparison in Fig. \ref{tomr} appears to confirm that it is larger in size than normal main sequence stars by a few percent. Together these 
characteristics appear to make star less luminous than model predictions by about 10\%. 

For the primary star, it appears to have a radius similar to other stars at the turnoff, but it is slightly hotter. To summarize the earlier discussion, the luminosity of DS And A appears to be significantly lower than
other well-studied eclipsing stars in the field and model predictions for the measured mass, and changes to chemical composition or physics (specifically convective core overshooting) are not capable of explaining the discrepancy. The temperature measured from the SED can be brought into better agreement with models if a higher bulk metallicity is adopted, although this appears to be contrary to what the cluster main sequence implies. The star's radius seems to be the characteristic that is least sensitive to metallicity uncertainties. Models can match the mass and radius of DS And A at the turnoff if core overshooting is at the high end of what is used in models currently ($\sim 0.3 H_P$, with the closest match being PARSEC models), but this does not fix the luminosity discrepancy.
 The characteristics of DS And A are more precisely determined than those of BD $+37$ 410, and they imply a much lower age: near
1.3 Gyr for PARSEC  and 
BaSTI-IAC isochrones, which are preferred based on convective overshooting arguments. However, we believe that the low age is another symptom of the peculiarities of DS And A. Although its CMD positions do not disagree wildly with those of other stars at the cluster turnoff, its mass does appear to be noticeably larger than would be expected based on the age derived from BD $+37$ 410. This probably points to it being an emerging blue straggler, and the age implied by the star probably should not be trusted for the cluster.

The future of the DS And binary system is fairly clear, based on its current configuration. The current primary star will evolve and expand to fill its Roche lobe relatively shortly (approximately 200 Myr), while it is still in the late stages of the main sequence. With the relatively slow mass transfer, the secondary star should be able to take in most of the mass of the secondary star without overflowing its Roche lobe, which would be expanding during the transfer. For a time the mass gainer would be a more obvious blue straggler until it evolved off the main sequence and onto the giant branch. Mass transfer at that point is likely to be unstable, leading to a common envelope phase and merger.

Based on the unusual present-day characterstics of the two stars, it is natural to wonder if the stars had an unusual formation history, or were somehow affected by evolving in the presence of another very nearby star.
Characteristics of the secondary star might be explained in part by magnetically-suppressed convective motions in surface layers. As discussed by \citet{clausen}, solar-type stars that are the lower-mass component in close binaries frequently appear to be older than the high-mass component, and there is evidence from X-ray emission that this is related to stellar activity driven by rapid rotation and enforced by tidal interaction. DS And is not the closest binary among systems showing these effects, nor is DS And B the most rapidly rotating of the stars. It is, however, more massive than stars previously identified, and presumably this means that its surface convection zone is significantly smaller in extent and less massive. The evidence may be enough to give it membership in the group, helped by the relatively short period. More examination of activity indicators for this system would be helpful. To date, the tentative detection of the binary in ROSAT PSPC data (a result of being outside the most reliable $20\arcmin$ of the observed field) by \citet{bvxray} remains the only X-ray observation.

DS And A, however, is too massive to have a significant surface convection zone on the main sequence, and so we should look elsewhere for an explanation of its characteristics. The stars of DS And appear to be relaxed and stable in their gravitational potentials, and so it is hard to imagine that the emitted luminosity of the primary could be lower than a star of the same mass would need to maintain its structure as a single star. A zero-age main sequence star should provide lower limits for both the luminosity and the radius, and we do find that DS And A is above both of these limits for a model star of its mass. 

The primary star appears somewhat evolved according to its radius, as if it had a different birth date than other cluster stars. A stellar merger might have resulted in a younger-looking star by bringing hydrogen-rich material into the core. However, if such a thing did occur, it must have happened early in the star's history or otherwise the remnant would not have had a chance to evolve as far away from the zero-age main sequence as it has. Based on the configuration of the binary now, such an event must have involved at least three stars, and would have had to be relatively long ago for the system to have circularized and synchronized. The low luminosity and temperature of the star is not explained in this kind of scenario though. Once a merger remnant relaxed, it should not have been much different than a normal single star --- the denser, lower-mass star in a merger would take up a place in the core of the star, giving it a lower core He abundance than a single star of its mass and the cluster's age would have. 

A binary mass transfer scenario has more substantial problems. Neither star in the binary appears to have enough mass to have evolved and expanded to reach its Roche lobe to initiate mass transfer. If the secondary star had originally been more massive and had initiated mass transfer in the distant past as a main sequence star, there is not a clear reason why mass transfer would terminate when the donor was a relatively normal-looking main sequence star of $1.18 \msun$.

While Li abundance measurements among bright NGC 752 main sequence stars are surprisingly few, the stars bracketing DS And A in magnitude (see Fig. \ref{rotmix}) have the highest abundances \citep{boesgaard}. Although measurement of the Li abundance of DS And A is beyond the scope of this paper, the lack of detectable Li in the primary star's atmosphere is a puzzle. 
If the star is the remnant of a merger early in the cluster's history, the more massive input star (that contributes the most to the remnant's surface layers) is unlikely to have depleted its photospheric Li greatly by the time of the merger. Even if it originated where the Li dip is seen in the cluster today, depletion probably would not have progressed significantly in a few hundred million years. This leaves the remnant's main sequence evolution as the most likely site of depletion.
Even though DS And A currently resides at the turnoff brighter than the Li dip where standard stellar evolution theory predicts that there should be no Li depletion, observations show that a fraction of stars are depleted and that this fraction increases as age increases \citep[e.g.,][]{deli6819}. For NGC 7789, with age similar to NGC 752, 13\% of the stars above the turnoff have Li abundances more than 1 dex away from the maximum for the cluster. The Li depletion appears to correlate with increasing spindown of cluster stars, so the enforced rapid rotation of DS And A within a tidally-locked binary might be expected to reduce depletion \citep{ryan}.

\subsection{Color-Magnitude Diagram (CMD) Comparisons and Cluster Age}

In light of the possible anomalies in the characteristics of the stars of 
DS And, it is worth re-examining other age indicators for the cluster.
There is a tremendous amount of photometric information available for the stars
of NGC 752 that can be employed for age dating the cluster turnoff.
Disagreements between models and observations for turnoff and subgiant stars
can reveal failings in the model physics, with core
convection being a particularly important consideration for stars in
this mass range. The extent of core convection has a pivotal role in
determining the hydrogen burning lifetime of the stars, and the way in
which stars leave the main sequence reveals details of the
distribution of hydrogen near the core. We will focus on PARSEC isochrones
because of the evidence that the overshooting algorithm they use is the best match for the characteristics of BD $+$37 410 and other single stars at the bright end of the main sequence.

We re-examined cluster membership based on Gaia Early Data Release 3
proper motions and parallaxes, and available radial velocity
information from \citet{agueros} and Gaia Data Release 2. We also examined
the wider field membership searches by \citet{bhatta} and \citet{boffin} in the hopes of 
collecting stars to fill in parts of the main sequence that are sparsely populated. 
Along the way we noticed an interesting clustering of radial velocities in the
8.5-10 \kms range, which includes several stars rejected as members by
\citeauthor{agueros} Examining this group of stars, we generally
found them to be consistent with membership based on proper motions,
parallaxes, and CMD position (frequently being very precisely on the
main sequence). We believe the evidence points toward membership for
these stars, although we don't have a good explanation for the
offset in radial velocity.

\begin{figure*}
\plotone{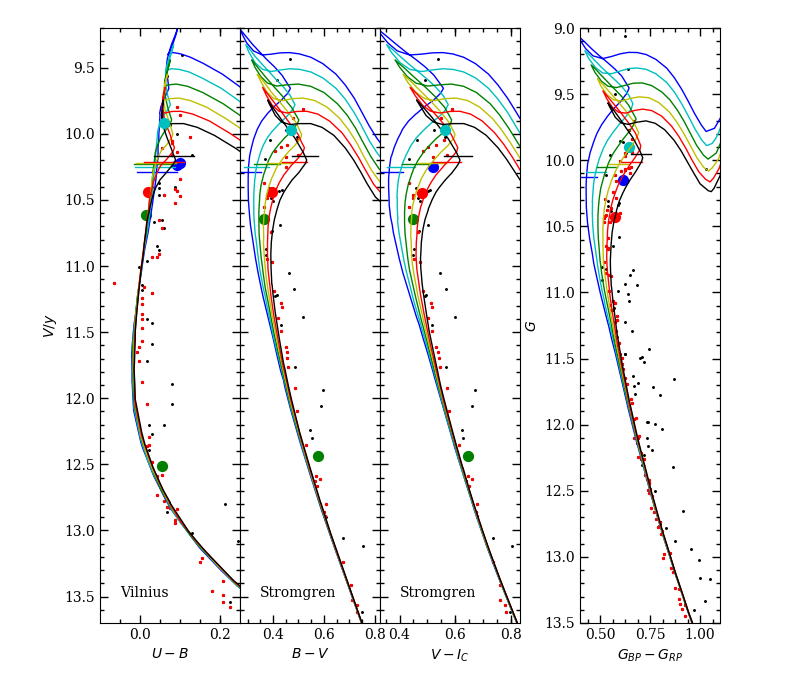}
\caption{Precise CMDs of likely single-star members of the NGC 752
  cluster (small red points) and binary members (black points).  Photometry of the DS And system at quadrature is shown
  with large red points, and the positions of the primary and secondary stars
  (as inferred from light curves) are shown with green points. Photometry of the BD $+$37 410 system is shown with cyan points, and a proxy for BD $+$37 410 A (PLA 641) is shown with blue points. PARSEC isochrones \citep{parsec} for $Z = 0.0152$
  are shown with ages from 1.3 (blue) to 1.8 (black) Gyr with 0.1 Gyr spacing. Horizontal lines mark the position of a star of BD $+37$ 410 A's mass on the isochrone.
\label{bestisos}}
\end{figure*}

We examined CMDs with the most precise photometry.
We used Vilnius photometry transformed to Johnson $U-B$ color, and Str\"{o}mgren photometry transformed to Johnson-Cousins $B-V$ and $V-I_C$ colors; see section \ref{sedsec}) in order to use high-precision photometry for cluster stars while making use of our photometric decomposition of $UBVR_CI_C$ for the DS And system using light curves. We show these CMDs, along with the Gaia EDR3 CMD  in Fig. \ref{bestisos}. While the ($U-B,V$) data and models disagree in several complex ways, the agreement is much better in other colors, and isochrone fits purely to the CMDs imply an age of $1.6-1.7$ Gyr. But beyond this, the models predictions are in good agreement with the measured photometry of the primary star within the uncertainties on its measured mass.

With the variety of photometry available for NGC 752 stars, we assembled an HR diagram for stars near the turnoff using SED fits and Gaia distance information.
We have generally restricted our sample to stars that have at least one measurement in $U$ or similar filters shortward of the Balmer decrement in order to gauge gravity effects on the SED.
The HR diagram is most consistent with solar metallicity --- if higher metallicity is used, the main sequence is predicted to be too red. The other factor that must be constrained is the amount of convective core overshooting. Redward extension of stars brighter than the turnoff is sensitive to the amount, with more overshooting prolonging the slow phase of core hydrogen burning and allowing the stars to grow bigger and more luminous before reaching core hydrogen exhaustion. As shown in Section \ref{mr}, the brightest turnoff stars support the larger amount of overshooting in the PARSEC models over the other isochrone sets. Thus, our preferred age for NGC 752 is near 1.6 Gyr, with an uncertainty of about 0.08 Gyr that comes from the scatter in the stars in the brightest main sequence.

\begin{figure}
  \plotone{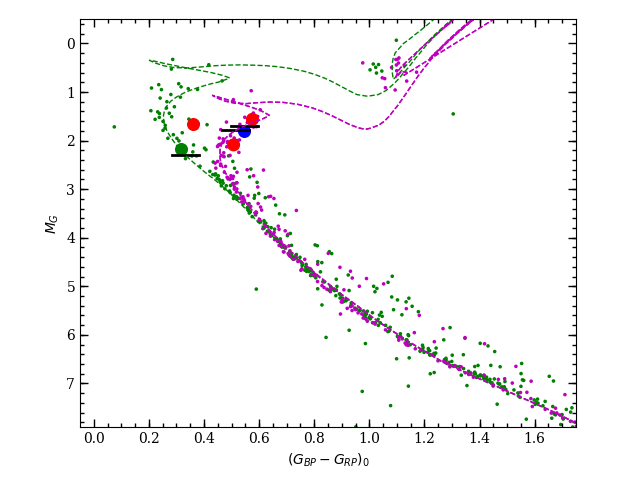}
 \caption{Color-magnitude diagram for NGC 752 (magenta points) and NGC
   6811 (green) cluster members selected based on {\it Gaia} proper
   motions and parallaxes from Early Data Release 3. PARSEC isochrones
   for $Z=0.0152$ and ages 1.00 and 1.60 Gyr are shown. The black marks
   shows where stars with the mass of KIC 9777062 A, DS And A, and BD $+$37 410 A would be in the respective
   isochrones. The large red points show the mean photometry for all three binaries,
   the large green point shows the approximate characteristics of
   KIC 9777062 A, and the blue point shows the characteristics of a proxy for BD $+$37 410 A.\label{abscmd}}
\end{figure}

We can derive ages relative to other clusters by selecting high probability members
based on {\it Gaia} proper motions and parallaxes, and using the parallaxes for distance corrections. 
Based on the apparent age of NGC 752, a good comparison
is NGC 6811, with an age close to 1 Gyr \citep{sand16}. NGC 6811 is close 
enough that parallaxes have also been measured for
main sequence stars far below the turnoff. Figure \ref{abscmd} shows a
comparison between likely members of the two clusters where the
photometry was corrected for the Gaia mean cluster parallaxes (2.24 mas and 0.87
mas for NGC 752 and NGC 6811, respectively) and for extinction
($E(B-V)=0.044$ and 0.07, respectively) using the prescription in
\citet{gaiacmd}. The main sequences of the two clusters overlap each
other extremely well except near the turnoff, clearly showing that NGC 752
is older. Using PARSEC models \citep{parsec}, we estimate that the two
clusters differ in age by about 0.6 Gyr.

The age that we prefer is larger than all recent measurements, and in some cases, significantly so.
An important factor in the discussion is the amount of convective core overshooting because the size of the mixed core during the main sequence phase affects the duration of core hydrogen burning while leaving most other observable characteristics nearly unchanged. So unless the
overshooting in the models is calibrated, there could be a substantial systematic error. We briefly discuss recent age measurements below.

\citet{barti} derived an age of $1.58\pm0.04$ Gyr, using Padova isochrones \citep{padova} to do a least-squares fit to Vilnius photometry for the cluster. Their procedure selected a model metallicity ($Z= 0.015$) close to what we use, but the distance modulus [$(m-M)_V=8.18\pm0.03$] that is somewhat smaller than implied by {\it Gaia} measurements. While their measurement is consistent with ours, Padova isochrones have larger overshooting ($\sim0.5 H_P$) than the successor PARSEC isochrones for stars in this mass range, but this is not likely to have affected their fits to main sequence stars a great deal.

\citet{twarog} found an age of $1.45\pm0.05$ Gyr using Victoria-Regina isochrones \citep{vandenberg2006} in their analysis of Str\"{o}mgren photometry. The Victoria-Regina isochrones have similar amounts of overshooting as PARSEC isochrones despite a different algorithm. The solar metal content used in the Victoria-Regina models ($Z_\odot = 0.0188$) is quite a bit larger than the PARSEC models ($Z_\odot = 0.0152$), although it is closer to the one recent re-examination of the solar metal content by \citet{magg}. Their best-fit models used [Fe/H]$=-0.04$ or $Z=0.0171$, so they had larger metal content than the models we used. If the model composition differences are taken into account, our estimations would probably be quite close. But this comparison emphasizes that uncertainties in the bulk metal content of the stars remains an important systematic.

Although \citet{gaiacmd} did not do a detailed exploration of parameter space, they were the
first to make use of {\it Gaia} photometry and distance information, alongside representative metallicity ([Fe/H]$=-0.03$) and reddening ($E(B-V)=0.04$) values. The authors derived an age of 1.4 Gyr using PARSEC isochrones, which we believe have an appropriate amount of convective overshooting, but they do also quote a fairly large uncertainty of more than 0.2 Gyr.

\citet{agueros} derived an age of $1.34\pm0.06$ Gyr for NGC 752 through
photometric SED fits to 59 of the most-likely single stars in their
membership survey. The cluster parameters were derived from the
combination of the individual posterior distributions for the single
star sample. While this is a statistically thorough age determination,
it is still subject to systematics. For example, their analysis only
employed MIST isochrones, which have an amount of overshooting that is too small to match the characteristics of the brightest main sequence stars. Only a portion of their sample had much
leverage on the age, as evidenced in their Figure 3 by a number of
stars that returned zero ages or ages two or more times the likely
age. In addition, their result for the cluster extinction $A_V$ was at
the high end of the prior range they searched, and is larger than
found in other studies. This may be a reflection of systematic
error as well.

\section{Conclusions}

An analysis of the eclipsing binary BD $+$37 410 reveals a primary star that has evolved past the cluster turnoff nearing the end of the heavily-populated part of the main sequence in NGC 752, and its characteristics clearly show the consequences of strong convective overshooting in the star's core that prolong the main sequence lifetime. The luminosity and radius of the star are inconsistent with predictions from some commonly-used sets of isochrones, and as a result, this observation greatly reduces systematic errors in the cluster's age associated with convective core overshooting. Overshooting algorithms are the subject of continuing research, and they are almost certain to be imperfectly representing the behavior of the convective core as a function of time in stars like those at the turnoff of NGC 752. But in terms of reproducing the extent of core convection shortly before core hydrogen exhaustion, PARSEC models are clearly coming closest. Based on CMD comparisons and especially the mass and radius of the primary star of BD $+$37 410, our preferred age is $1.61\pm0.03\pm0.05$ Gyr. 

We have also presented measurements of the close eclipsing binary system DS And, which contains a
star near the turnoff of the nearby open cluster NGC 752. We have improved the precision and accuracy of the radial velocity measurements for the binary, which lead to precise masses $M_1 = 1.692\pm0.004\pm0.010 \msun$ and $M_2 = 1.184\pm0.001\pm0.003 \msun$. Light curves in nine different wavelength bands allow us to separate the contributions to the system light from the two stars and to derive effective temperatures. Analysis of the eclipsing light curves allows us to derive precise radii $R_1 = 2.185\pm0.004\pm0.008 \rsun$ and $R_2 = 1.200\pm0.003\pm0.005 \rsun$.
The stars are in a very close binary, although they are pretty clearly detached, with a few percent distortion of the shape of the primary star. The stars rotate significantly faster than stars of similar brightness in the cluster: $v_{rot} \sin i = 105, 58$ \kms, while likely single stars in the cluster generally have $v_{rot} < 35$ \kms \citep{mmu}.  A relatively high amount of convective core overshooting appears to be needed to match its radius at its evolutionary stage at the turnoff. The star's temperature is in line with other stars at the turnoff, and the star's radius is clearly lower than expected when we compare to the primary star of BD $+$37 410. The star's luminosity is also significantly lower than predicted for reasonable models. We do not have a clear explanation for all of the primary star's characteristics, but a stellar merger early in the cluster's history is consistent with the most data.

Together, both binary systems clearly show that stars with mass near $1.7 \msun$ inhabit NGC 752's turnoff and are in the last stages of main sequence evolution. The best-fitting models predict that stars with masses of $1.77-1.80 \msun$ are leaving the main sequence. This is important to know because of the well-known fact that NGC 752 is one of the few star clusters known with red clump stars covering a large range of magnitudes, indicating its most massive stars are making the transition from non-degenerate helium ignition (which happens at relatively low He core mass, and results in lower-luminosity clump stars) to degenerate He flash ignition. While a detailed analysis of the clump stars is beyond our scope here, the mass information should be very valuable in further examination of the physics governing the giant branch and helium ignition at this transition.

\begin{acknowledgments}
E.L.S. gratefully acknowledges support from the National Science Foundation 
under grant AAG 1817217. We thank the anonymous referee for a thorough reading of the manuscript and for many helpful comments on presentation. We thank K. Brogaard for providing the
original version of the spectral disentangling code used in this work,
B. Twarog for providing his Str\"{o}mgren photometry of DS And, and H. Boffin for providing the
results of his membership study for NGC 752.  We
would also like to thank D. Baer, E. Bavarsad, D. Jaimes, and
M. J. Jeffries, Jr. for assisting in the acquisition of ground-based
photometric observations.

This research made use of observations from the SIMBAD database,
operated at CDS, Strasbourg, France; and the WEBDA database, operated
at the Institute for Astronomy of the University of Vienna; data
products from the AAVSO Photometric All Sky Survey (APASS), funded by
the Robert Martin Ayers Sciences Fund and the National Science
Foundation.
The Hobby–Eberly Telescope (HET) is a joint project of the University
of Texas at Austin, the Pennsylvania State University, Stanford
University, Ludwig-Maximilians-Universit\"{a}t M\"{u}nchen, and
Georg-August-Universit\"{a}t G\"{o}ttingen. The HET is named in honor of its
principal benefactors, William P. Hobby and Robert E. Eberly.
\end{acknowledgments}

\facilities{MLO:1m, HET (HRS), NOT (FIES), WIYN (Hydra), TESS}

\software{IRAF \citep{iraf1,iraf2}, DAOPHOT \citep{daophot}, ELC \citep{elc}, eleanor \citep{eleanor}, FIEStool ({\tt http://www.not.iac.es/instruments/fies/fiestool/FIEStool.html}), QLP \citep{qlp}, BF-rvplotter ({\tt https://github.com/mrawls/BF-rvplotter})}

\appendix

\section{Photometry Sources and Spectral Energy Distribution Fitting}\label{photapp}

Table \ref{photsource} holds information on the photometry we used to delineate the SEDs of NGC 752 stars.
We provide references for the photometry and for the calibration to flux.

\begin{deluxetable}{lclllc}
\tabletypesize{\scriptsize}
\tablewidth{0pc}
\tablecaption{Sources of Photometry for NGC 752 Spectral Energy Distributions}
\tablehead{
\colhead{Facility/Survey} & \colhead{Filter} & \colhead{$\lambda_{eff}$ (\AA)} & \colhead{References} & \colhead{Calibration} & \colhead{Notes}}
\startdata
GALEX & FUV    & 1538.6 & \citet{galex} & \citet{galexcal}\\
      & NUV    & 2315.7 & \\
Swift-UVOT & $uvw2$ & 2030   & \citet{siegel}\\
           & $uvm2$ & 2231   &\\
           & $uvw1$ & 2634   & \\
XMM-Newton & $uvw1$ & 2971 & & & 1\\
Vilnius & $U_V$  & 3450   & \citet{barti} & \citet{mvb}\\
        & $P_V$  & 3740   & \citet{zdana}\\
        & $X_V$  & 4054   & \\
        & $Y_V$  & 4665   & \\
        & $Z_V$  & 5162   & \\
        & $V_V$  & 5442   & \\
        & $S_V$  & 6534   & \\
Str\"{o}mgren & $u$    & 3520   & \citet{twarog} & \citet{gray-strom}\\
              & $v$    & 4100   & \citet{hmstrom}\\
              & $b$    & 4688   & \\
              & $y$    & 5480   & \\
Geneva & $U_G$ & 3471 & \citet{rufener} & & 2\\
       & $B1_G$ & 4023 & \\
       & $B_G$ & 4246 & \\
       & $B2_G$ & 4482 & \\
       & $V1_G$ & 5402 & \\
       & $V_G$ & 5504 & \\
       & $G_G$ & 5814 & \\
$UBV$ Means & $U$    & 3663   & \citet{mmubv} & \citet{bcp} & 3\\
            & $B$    & 4361   & \\
            & $V$    & 5448   & \\
Tycho & $B_T$  & 4220   & \citet{tycho} & \citet{mvb} & 6\\
      & $V_T$  & 5350   & \\
D94 & $B$    & 4361   & \citet{daniel} & \citet{bcp} & 3\\
    & $V$    & 5448   & \\
APASS & $B$    & 4361   & \citet{apass} & \citet{bcp} & 3,9\\
      & $V$    & 5448   & \\
      & $g^\prime$ & 4640 & \\
      & $r^\prime$ & 6122 & \\
      & $i^\prime$ & 7440 &\\
Pan-STARRS1 & $g_{P1}$ & 4810  & \citet{ps1desc} & \citet{ps1} & 5\\
            & $r_{P1}$ & 6170  & \\
            & $i_{P1}$ & 7520  & \\
            & $z_{P1}$ & 8660  & \\
            & $y_{P1}$ & 9620  & \\
T08 & $V$ & 5448 & \citet{taylorvri} & \citet{bcp} & 3\\
     & $R_C$ & 6414   &  \\
     & $I_C$ & 7980   &  \\
TASS & $V$ & 5448 & \citet{tass} & \citet{bcp} & 3,4\\
     & $I_C$ & 7980   &  \\
Gaia & $G_{BP}$ & 5051.5 & & & 7\\
     & $G$    & 6230.6 & \\
     & $G_{RP}$ & 7726.2 & \\
2MASS & $J$   & 12350   & \citet{2mass} & \citet{2masscal} & 8\\
      & $H$   & 16620   &  \\
      & $K_s$ & 21590   &  \\
WISE & $W1$  & 33526   & \citet{wise} & \\
     & $W2$  & 46028   &  \\
     & $W3$  & 115608  &  \\
\enddata
\label{photsource}
\tablecomments{1) {\it XMM-Newton} Optical Monitor Serendipitous Source Catalogue (Version 4.1).
2) Retrieved from WEBDA. 3) Measurements on a Vega magnitude
system, converted to fluxes using reference
magnitudes from Table A2 of \citet{bcp}, accounting for the
reversal of the zero point correction rows for $f_\lambda$ and
$f_\nu$. 4) The Amateur Sky Survey Mark IV catalog version 2. 5) Mean PSF magnitudes used. 6) Tycho-2 Catalogue. 7) Early Data Release 3. 8) Photometry from the All-Sky Point Source Catalog. 9) Data Release 10.}
\end{deluxetable}

The {\it Galaxy Evolution Explorer} ({\it GALEX}) observations of the cluster involved guest investigator images in the FUV for 4539 s (GI1 proposal 27, P.I. K. Honeycutt),
and NUV exposures as part of the All-Sky Imaging Survey
(AIS). Archived magnitudes are based on count rates with minimal
background contributions, so we computed the magnitude and flux based
on the average count rate for redundant AIS observations in the NUV.
\citet{galexcal} describes the characteristics of the GALEX photometry
and its calibration to flux. 

We fitted the photometric data for $\teff$ and bolometric flux
$F_{bol}$ using ATLAS9 models \citep{atlas9}\footnote{The models were
  calculated using the ATLAS9 fortran code that employed updated 2015
  linelists, and we computed models at finer temperature intervals
  than are available in the published grid.}.
Models were adjusted to account for the interstellar reddening of the
cluster ($E(B-V)=0.044\pm0.0034$; \citealt{taylor}) using the
\citet{cardelli} extinction curve. We commonly have flux information from the ultraviolet into
the infrared, so that $\teff$ is delimited well. However, we employed the
infrared flux method \citep[IRFM;][]{irfm} to get better precision on
the temperature. The ratio of bolometric and infrared fluxes (in
2MASS bands) can be compared with an approximate ATLAS9 model to compute a new estimate of $\teff$.
We calculated $\teff$ values from all three 2MASS bandpasses, and we used the average
as the most representative value. We used the 2MASS flux calibration of \citet{casairfm}
because it produced better consistency between the temperatures
from the three bands than did the \citet{2masscal} calibration.
A new ATLAS9 model was computed using the updated $\teff$ and the process was iterated until $\teff$ in the model and $\teff$ from the IRFM agreed.

For cluster single stars, we generally only fitted stars with at least one measurement in $U$ or similar bandpasses in order to monitor whether the gravity-sensitive Balmer decrement feature was being modelled properly. Trends in the flux residuals ($\Delta F_\lambda / F_\lambda$) as a function of wavelength were monitored to ensure that there were not gross errors in the temperatures. Clearly discrepant photometric values were removed before the fits were complete.

\begin{deluxetable}{lrcrcrc}
\tablewidth{0pt}
\tabletypesize{\scriptsize}
\tablecaption{Radial Velocity Measurements}
\tablehead{\colhead{mJD\tablenotemark{a}} & \colhead{$v_A$} & \colhead{$\sigma_{A}$} & \colhead{$v_B$} & \colhead{$\sigma_B$} & \colhead{$v_C$} & \colhead{$\sigma_C$}\\
& \multicolumn{2}{c}{(\kms)} & \multicolumn{2}{c}{(\kms)} & \multicolumn{2}{c}{(\kms)}}
\startdata
\multicolumn{5}{c}{DS And} \\
\multicolumn{5}{c}{NOT FIES Observations} \\
54335.55788 &  125.15 & 1.18 &$-167.19$& 1.03 \\
54336.53919 &  125.54 & 1.20 &$-172.07$& 1.78 \\
54336.60635 &  117.35 & 1.30 &$-152.54$& 1.12 \\
54336.75052 &   44.80 & 1.31 & $-60.97$& 1.99 \\
54337.49382 &  118.94 & 1.11 &$-159.96$& 1.14 \\
54337.56653 &  127.33 & 1.47 &$-170.93$& 1.18 \\
54671.59838 &$-105.79$& 0.90 &  166.21 & 1.64  \\
54671.62437 & $-96.08$& 0.63 &  148.69 & 1.31 \\
54671.67465 & $-68.85$& 0.73 &  109.15 & 1.05 \\
54693.60647 & $-67.93$& 0.86 &  101.64 & 0.65 \\
54696.64674 & $-72.44$& 0.54 &  114.02 & 1.12 \\
54707.53952 &   87.36 & 0.89 &$-104.93$& 1.36 \\
54762.43073 &$-112.28$& 0.71 &  176.64 & 0.49 \\
54762.49913 &$-114.01$& 1.03 &  181.60 & 1.27 \\
54787.63134 & $-90.00$& 0.75 &  143.26 & 1.30 \\
\multicolumn{5}{c}{HET HRS Observations} \\
55420.87490 &  126.50 &  1.18 & $-162.40$&  1.80 \\
55421.86256 &  128.09 &  0.98 & $-169.50$&  1.31 \\
55466.98209 & $-57.56$&  2.05 &    95.77 &  2.36 \\
55466.98701 & $-51.80$&  2.28 &    92.64 &  1.45 \\
55470.73432 & $-88.30$&  1.35 &   132.36 &  0.50 \\
55476.94538 &$-115.92$&  1.40 &   178.68 &  1.55 \\
55478.94247 &$-118.67$&  1.33 &   182.22 &  2.54 \\
55479.92220 &$-116.22$&  1.24 &   180.50 &  2.59 \\
55486.92143 & $-98.73$&  1.25 &   143.91 &  1.66 \\
55493.89688\tablenotemark{b} & $-42.86$&  1.56 &   82.95 &  0.65 \\
55505.87386 &   73.52 &  0.65 &  $-88.84$&  1.30 \\
\enddata
\label{spectab}
\tablenotetext{a}{mJD = BJD - 2400000.}
\tablenotetext{b}{Not used in fit.}
\tablecomments{Table \ref{spectab} is published in its entirety in the
machine-readable format. A portion is shown here for guidance regarding
its form and content.}
\end{deluxetable}

\begin{deluxetable}{lc|lll|lll|l}
\tabletypesize{\scriptsize}
\tablewidth{0pc}
\tablecaption{Photometry of DS And and BD $+37$ 410}
\tablehead{
\colhead{Filter} & $\lambda_{eff}$ & \colhead{$m_\lambda$} & \colhead{$\sigma_m$} & \colhead{$F_\lambda$} 
& \colhead{$m_\lambda$} & \colhead{$\sigma_m$} & \colhead{$F_\lambda$} & \colhead{Notes}\\
 & \colhead{(\AA)} & & & \colhead{(erg / cm$^2$ s \AA)} & & & \colhead{(erg / cm$^2$ s \AA)} &}
\startdata
FUV    & 1538.6 & 18.394 & 0.0137 & $2.073\times 10^{-15}$ & 19.0833 & 0.0209 & $1.099\times 10^{-15}$ & \\
NUV    & 2315.7 & 14.599 & 0.0079 & $3.208\times 10^{-14}$ & 14.2347 & 0.0014 & $4.487\times 10^{-14}$ & \\ 
$U_V$  & 3450   & 12.704 &       & $1.554\times 10^{-13}$ & 12.260 &  & $2.339\times 10^{-13}$ & 2\\
$U_G$  & 3471   &        &       &                        & 10.955 &  & $2.388\times 10^{-13}$ & 10\\
$u$    & 3520   & 12.245 & 0.022 & $1.482\times 10^{-13}$ & 11.711 & 0.004 & $2.424\times 10^{-13}$ & 1\\
$U$    & 3663   & 10.852 & 0.005 & $1.905\times 10^{-13}$ & & & & 7\\
$U$    & 3663   & 10.915 & 0.009 & $1.798\times 10^{-13}$ & 10.473 & 0.004 & $2.701\times 10^{-13}$ & 4\\
$P_V$  & 3740   & 12.177 &       & $1.925\times 10^{-13}$ & 11.71 &  & $2.960\times 10^{-13}$ & 8,2 \\
$B1_G$ & 4023   &        &       &                        & 10.584 & & $3.948\times 10^{-13}$ & 10\\
$X_V$  & 4054   & 11.541 &       & $2.728\times 10^{-13}$ & 11.09 &  & $4.133\times 10^{-13}$ & 8,2\\
$v$    & 4100   & 11.210 & 0.018 & $2.841\times 10^{-13}$ & 10.705 & 0.005 & $4.524\times 10^{-13}$ & 1\\
$B_T$  & 4220   & 11.140 & 0.052 & $2.379\times 10^{-13}$ & 10.431 & 0.035 & $4.571\times 10^{-13}$ & \\
$B_G$  & 4246   &        &       &                        & 9.583 & & $4.234\times 10^{-13}$ & 10\\
$B$    & 4361   & 10.833 & 0.009 & $2.935\times 10^{-13}$ & & & & 7\\
$B$    & 4361   & 10.913 & 0.003 & $2.726\times 10^{-13}$ & 10.438 & 0.001 & $4.223\times 10^{-13}$ & 5\\
$B$    & 4361   & 10.89  &       & $2.785\times 10^{-13}$ & 10.435 & 0.004 & $4.234\times 10^{-13}$ & 4\\
$B$    & 4361   & 10.862 & 0.061 & $2.858\times 10^{-13}$ & 10.517 & 0.182 & $3.926\times 10^{-13}$ & 3\\
$B2_G$ & 4482   &        &       &                        & 10.952 & & $4.512\times 10^{-13}$ & 10\\
$Y_V$  & 4665   & 10.922 &       & $2.905\times 10^{-13}$ & 10.44 &  & $4.529\times 10^{-13}$ & 2\\
$g^\prime$ & 4640    & 10.650 & 0.062 & $2.775\times 10^{-13}$ & 10.355 & 0.133 & $3.640\times 10^{-13}$ & 3 \\
$b$    & 4688   & 10.770 & 0.012 & $2.898\times 10^{-13}$ & 10.255 & 0.003 & $4.657\times10^{-13}$ & 1\\
$g_{P1}$ & 4810   &   &  & & 10.038 & 0.010 & $4.537\times 10^{-13}$ & \\
$G_{BP}$ & 5051.5 & 10.7285 & 0.0396 & $2.120\times 10^{-13}$ & 10.0748 & 0.0029 & $3.871\times 10^{-13}$ & \\
$Z_V$  & 5162   & 10.668 &       & $2.477\times 10^{-13}$ & 10.12 &  & $4.103\times 10^{-13}$ & 8,2\\
$V_T$  & 5350   & 10.625 & 0.048 & $2.266\times 10^{-13}$ & 9.898 & 0.031 & $4.426\times 10^{-13}$ & \\
$V1_G$ & 5402   &        &       &                        & 10.700 & & $3.916\times 10^{-13}$ & 10\\
$V_V$  & 5442   & 10.504 &       & $2.371\times 10^{-13}$ & 9.95 &  & $3.950\times 10^{-13}$ & 8,2\\
$V$    & 5448   & 10.439 & 0.004 & $2.423\times 10^{-13}$ & 9.967 & 0.047 & $3.743\times 10^{-13}$ & 7,9\\
$V$    & 5448   & 10.506 & 0.037 & $2.278\times 10^{-13}$ & 9.968 & 0.013 & $3.739\times 10^{-13}$ & 5\\
$V$    & 5448   & 10.480 & 0.017 & $2.333\times 10^{-13}$ & 9.963 & 0.030 & $3.757\times 10^{-13}$ & 4\\
$V$    & 5448   & 10.439 & 0.061 & $2.423\times 10^{-13}$ & 10.036 & 0.003 & $3.512\times 10^{-13}$ & 3\\
$y$    & 5480   & 10.510 & 0.010 & $2.332\times 10^{-13}$ & 9.944 & 0.003 & $3.927\times 10^{-13}$ & 1\\
$V_G$  & 5504   &        &       &                        & 9.963 & & $3.865\times 10^{-13}$ & 10\\
$G_G$  & 5814   &        &       &                        & 11.037 & & $3.651\times 10^{-13}$ & 10\\
$r^\prime$ & 6122 & 10.404 & 0.070 & $1.999\times 10^{-13}$ & & & & 3\\
$r_{P1}$ & 6170  &          &     &  & 9.836 & 0.010 & $3.321\times 10^{-13}$ &  \\
$R_C$  & 6414   & 10.222  & 0.005 & $1.775\times 10^{-13}$ & & & & 7\\
$G$    & 6230.6 & 10.4555 & 0.0088 & $1.671\times 10^{-13}$ & 9.9012 & 0.0031 & $2.784\times 10^{-13}$ & \\
$S_V$  & 6534   & 10.084  &       & $1.724\times 10^{-13}$ & 9.48 &  & $3.008\times 10^{-13}$ & 8,2\\
$i^\prime$ & 7440 & 10.341 & 0.058 & $1.435\times 10^{-13}$ & & & & 3\\
$i_{P1}$ & 7520  &  &       &  & 9.786 & 0.079 & $2.341\times 10^{-13}$ & \\
$G_{RP}$ & 7726.2 & 10.1141 & 0.0285 & $1.183\times 10^{-13}$ & 9.4316 & 0.0039 & $2.218\times 10^{-13}$ & \\
$I_C$  & 7980   &  9.959 & 0.012 & $1.169\times 10^{-13}$ & 9.407 & 0.066 & $1.944\times 10^{-13}$ & 7,9\\
$z_{P1}$ & 8660  & & & & 9.784 & 0.010 & $1.769\times 10^{-13}$ & \\
$y_{P1}$ & 9620  & & & & 9.784 & 0.010 & $1.433\times 10^{-13}$ & \\
$J$   & 12350   &  9.653 & 0.018 & $4.307\times 10^{-14}$ & 9.001 & 0.023 & $7.852\times 10^{-14}$ & \\
$H$   & 16620   &  9.481 & 0.019 & $1.827\times 10^{-14}$ & 8.792 & 0.018 & $3.447\times 10^{-14}$ & \\
$K_s$ & 21590   &  9.407 & 0.014 & $7.395\times 10^{-15}$ & 8.696 & 0.017 & $1.423\times 10^{-14}$ & \\
$W1$  & 33526   &  9.352 & 0.023 & $1.486\times 10^{-15}$ & 8.695 & 0.023 & $2.721\times 10^{-15}$ & 6\\
$W2$  & 46028   &  9.368 & 0.019 & $4.322\times 10^{-16}$ & 8.724 & 0.021 & $7.822\times 10^{-16}$ & 6\\
$W3$  & 115608  &  9.383 & 0.031 & $1.150\times 10^{-17}$ & 8.720 & 0.022 & $2.118\times 10^{-17}$ & \\
\hline
\enddata
\label{sedtable}
\tablerefs{1: \citet{twarog}. 2. \citet{zdana} 3. APASS (v. 10). 4. \citet{mmubv}
5. \citet{daniel} 6. \citet{chen18} measurement at light curve maximum.
7. \citet{milone19} measurement at light curve maximum. 8. \citet{barti}. 9. \citet{tass}. 10. \citet{rufener}.}
\end{deluxetable}

\end{document}